\documentclass[structabstract]{aa}  
\usepackage{graphicx}
\usepackage{natbib}

\begin{document}

   \title{The star cluster - field star connection in nearby spiral galaxies}
   \subtitle{ I. Data analysis techniques and application to NGC 4395}
 
   \author{E. Silva-Villa
          \and
          S. S. Larsen
          }

   \institute{Astronomy Institute, University of Utrecht,
              Princetonplein 5, 3584 CC, Utrecht, The Netherlands\\
              \email{[e.silvavilla,s.s.larsen]@uu.nl}
         			}
 							

  \abstract 
	{It is generally assumed that a large fraction of stars are initially born in clusters. However,
	a large fraction of these disrupt on short timescales and the stars end up belonging to the
	field. Understanding this
	process is of paramount importance if we wish to constrain the star formation histories
	of external galaxies using star clusters.} 
	{We attempt to understand the relation between field stars and star 
	clusters by simultaneously studying both in a number of nearby galaxies. } 
	{As a pilot study, we present results for the late-type spiral NGC 4395 
	using HST/ACS and HST/WFPC2 images. Different detection criteria were used
	to distinguish point sources (star candidates) and extended objects (star cluster candidates). 
	Using a synthetic CMD method, we estimated the star formation history. 
	Using simple stellar population model fitting, we calculated the mass and age of the cluster candidates.}
	{The field star formation rate appears to have been roughly constant, or to have possibly increased by up 
	to about a factor of two, for ages younger than $\sim$300 Myr within the 
	fields covered by our data. Our data do not allow us to constrain the star formation histories at older ages. We 
	identify a small number of clusters in both fields. Neither massive ($>10^5$ M$_\odot$) 
	clusters nor clusters with ages $\geq1$ Gyr were found in the galaxy and we found few clusters older than 100 Myr.} 
	{Based on our direct comparison of field stars and clusters in NGC~4395,
	we estimate the ratio of star formation rate in clusters that survive for $10^7$ to $10^8$
	years to the total star formation to be $\Gamma\sim0.03$. We suggest that this relatively low
	$\Gamma$ value is caused by the low star formation rate of NGC~4395.}
 

   \maketitle
%


\section{Introduction}
It is commonly assumed that most (if not all) stars are formed
in clusters. Clusters, due to dynamical and stellar evolution,
dissolve \citep{spitzer87} and the stars that belong to them become
part of the field stellar population of the galaxy. 
To use clusters as effective tools to constrain the formation and
evolution of galaxies, it is necessary to improve our understanding of
what fraction of stars end up as members of bound clusters and in the
field, respectively.

Both field stars and star clusters have been studied extensively 
in the Milky Way and nearby galaxies. In principle, field stars hold information
about star formation histories over the entire Hubble time and can therefore provide
important information about galaxy formation and evolution, e.g., \citet[][]{edvardsson93}.
Main sequence stars are present at all ages, and observations of the main sequence
turn-off can constrain epochs of stellar formation especially in dwarf galaxies where
relatively distinct bursts are often observed \citep{mateo98}.
Other features of the color-magnitude diagram (CMD), such as the sub-giant and
horizontal branches, giant and asymptotic branches, red clump stars, and 
red and blue supergiants can provide information about specific epochs of star formation.
However, it is necessary to apply a more sophisticated modeling of
the CMD to reconstruct star formation histories, taking into account different 
effects, e.g., incompleteness, resolution, depth of the observations, extinction and chemical
composition. \citet{tosi91} developed a method that takes these effects into account and attempts to use all the information
available in a CMD to reconstruct the field star formation history of a galaxy. The synthetic CMD method 
creates a synthetic population that is compared to the observed CMD
to constrain the star formation history (SFH) of the field stars in a galaxy. Many subsequent
studies have refined this method \citep{dolphin97,harriszaritsky01,coimbra02}. The SFHs of many galaxies in the Local Group and nearby have been studied using the synthetic CMD method:
\citet{brown08} for M31, \citet{harriszaritsky09} for the LMC, \citet{harriszaritsky04} for the SMC, \citet{barker07} for the M33,
\citet{cole07} for Leo A, \citet{young07} for Phoenix, \citet{annibali09} for NGC 1705, \citet{williams09} 
for M81, \citet{larsen07} for NGC1313 and \citet{rejkuba04} for NGC 5128, among many others.

On the other hand, the study of cluster systems and disruption processes can provide
important insight into the origin of field stars in a galaxy. Clusters disrupt by means of a variety of
mechanisms, including ``infant mortality'', stellar evolution, two-body relaxation, and tidal shocks that have been extensively 
studied by \citet{BL03,LL03,lamers05,baumgardt09,fall06,elmegreen08,whitmore07}, and \citet{bastiangieles08}
among others. The analysis of star clusters is often based on a comparison between the
 observed spectral energy distribution and theoretical models, which provides information about the ages and masses of the studied clusters and,
in turn, their dissolution. However, the comparison of models with observations also remains
affected by many uncertainties, e.g., binarity, completeness effects.

Some studies have started to address the relation between field stars and
clusters more explicitly. The number of stars that were formed in clusters has been estimated
to be $70\%-90\%$ in the solar neighbourhood, while $50\%-95\%$ of these embedded
clusters dissolve in a few Myrs \citep{LL03,lamersgieles08}. 
\citet{gielesbastian08} estimated that only 2-4\% of the global star formation rate in the SMC happened in bound star clusters. It is currently unknown what fraction of stars are initially born in clusters in SMC, but if this fraction were as large as in the Solar neighbourhood this would imply that there is also a large infant mortality rate in the SMC.
\citet{bastian08} studied the relation between the cluster formation rate and the star formation rate ($\Gamma=\frac{CFR}{SFR}$)
using archival Hubble images for high star formation rate galaxies and additional galaxies/clusters from literature. He found this fraction to be $\Gamma\sim0.08$ and thus concluded that the fraction of stars
formed in (bound) clusters represents $8\%$ of the total star formation. On the other hand,
\citet{gieles09} found that $\Gamma$ is given by $0.05\leq\Gamma\leq0.18$ in the galaxies M74, M101, and M51.
In most of these cases, however, it is difficult to tell whether there is a genuine ``field'' mode of cluster formation, or whether \emph{all}
stars form initially in clusters of which a large fraction dissolves rapidly. 

We aim to analyze the relation between field stars and star clusters
in different environments and address the question of whether or not there is a
constant cluster formation ``efficiency''. To this end, we use \textit{Hubble Space Telescope} (HST)
images of a set of five galaxies  (NGC 4395, NGC 1313, NGC 45, NGC 5236, and NGC 7793), which are nearby, face-on, spirals that 
differ in their morphologies, star, and cluster formation histories.
The cluster systems of these galaxies were studied by \citet{mora09} using the same Hubble images analyzed in this work. Mora et al. observed significant variations in the cluster age distributions of these galaxies. The galaxies are sufficiently nearby ($\sim4$ Mpc) for the brighter field stars to be well resolved in HST images, so that (recent) field star formation histories can be constrained by means of the synthetic CMD method. We can therefore take advantage of the superb spatial resolution of HST images to study field stars and clusters simultaneously within specific regions of these galaxies.

As a pilot work, this paper is devoted to presenting and testing all the analysis procedures, such as detection of stars and cluster candidates, completeness tests and photometry, and derivation of field star and cluster age distributions. We discuss our implementation of the synthetic CMD method as an IDL program and carry out tests of this program. 
To test our procedures we use the galaxy NGC 4395. The methods described in this paper will be used in our study of the rest of the galaxy sample (Silva-Villa et al.\ 2010, in prep.).
 
This article has the following structure. In Sect. 2, we provide general
information about NGC~4395. We present the
observations, data reduction, and photometry, where we differentiate the
point sources (star candidates) from the extended objects (cluster
candidates) in Sects. 3 and 4. Section 5 is devoted to
the analysis of the star and cluster properties. Finally, in Sect.
6 we present the discussion and conclusions.

\begin{table*}[!t]\centering
  \newcommand{\DS}{\hspace{6\tabcolsep}} 
  \caption{Journal of HST/ACS and HST/WFPC2 observations for both fields (F1 and F2) in NGC 4395.}
  \begin{tabular}{c @{\DS} cccc }
    Proposal ID & Date & Filter & Total exp. time (s) \\ \hline \hline
    \multicolumn{3}{c}{NGC 4395 (J2000.0) $\alpha:12^h 26^m 00^s$ $\delta: +33^o 31^{'} 04^"$}\\   
    9774 & 2004 Jun 12 & F435W (B) & 680 \\
    9774 & 2004 Jun 12 & F555W (V) & 680 \\
    9774 & 2004 Jun12 & F814W (I) & 430 \\ \hline
    \multicolumn{3}{c}{NGC 4395 (J2000.0) $\alpha:12^h 26^m 00^s.50$ $\delta: +33^o 30^{'} 58^".3$}\\
    9774 & 2004 Jun 12 & F336W (U) & 2400 \\ \hline
    \multicolumn{3}{c}{NGC 4395 (J2000.0) $\alpha:12^h 25^m 45^s.20$ $\delta: +33^o 34^{'} 28^"$}\\ 
    9774 & 2004 Jun 18 & F435W (B) & 680 \\
    9774 & 2004 Jun 18 & F555W (V) & 680 \\
    9774 & 2004 Jun 18 & F814W (I) & 430 \\ \hline
    \multicolumn{3}{c}{NGC 4395 (J2000.0) $\alpha:12^h 25^m 42^s.72$ $\delta: +33^o 34^{'} 22^".6$}\\ 
    9774 & 2004 Jun 18 & F336W (U) & 2400 \\ \hline \hline
  \end{tabular}
  \label{tab:data}
\end{table*}

\section {NGC 4395}

According to the NASA/IPAC Extragalactic Database (NED), NGC~4395 is a late-type spiral classified as type SA(s)m. 
It harbours the closest and least luminous known example of a Seyfert 1 nucleus \citep{filippenkosargent89}. Following \citet{LR99}, we adopt a distance modulus of $(m-M)=28.1 \ (D\approx4.2$ Mpc) and an absolute magnitude $M_B = -17.47$, intermediate between the Small and Large Magellanic Cloud. We assume a Galactic foreground extinction for NGC~4395 of $A_B=0.074$ \citep{schlegel98}.

\begin{figure}[!h]
	\centering
		\includegraphics[width=\columnwidth]{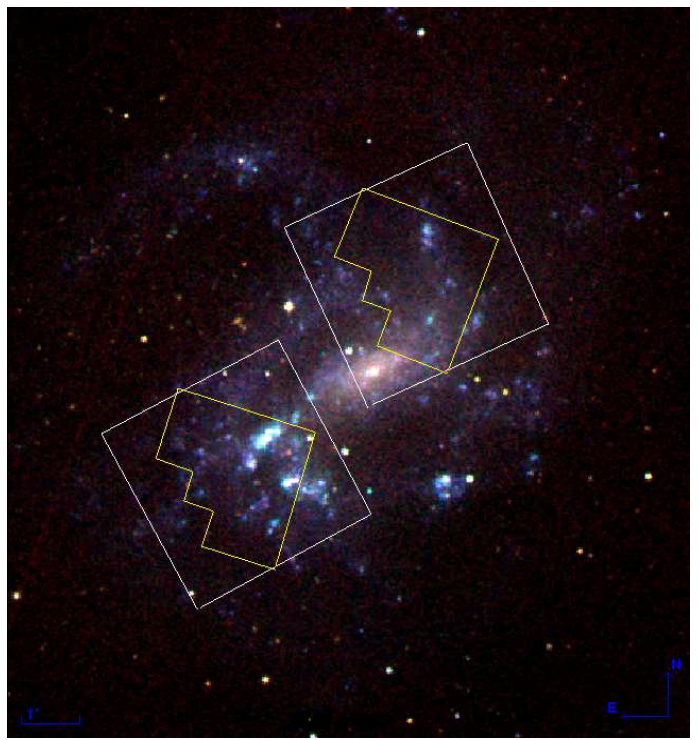}
	\caption{Second Palomar Sky Survey images of NGC 4395 combined using Aladin software. HST/ACS (white lines) and HST/WFPC2 
		     (yellow lines) fields covered by our observations are indicated.}
	\label{fig:ngc4395}
\end{figure}

The cluster system of NGC~4395 was first studied by \citet{LR99}, using
ground-based imaging. Using multiband ($UBVRIH\alpha$) photometry, Larsen \& Richtler identified 2 young clusters in this galaxy, although their observations were limited to objects with $M_V\leq -8.5$.
In their work, NGC 4395 is part of a sample of 21 galaxies. Compared to the remaining sample, 
NGC 4395 exhibits an exceptionally small number of star clusters and a low specific luminosity 
\citep[ratio of cluster- to total galaxy light;][]{LR00}.
Using HST images, \citet{mora09} could detect clusters based on their sizes and found a total of
44 clusters in NGC~4395 to a magnitude limit of $M_B = -3.3$. Compared to the other 4 galaxies
in their sample, Mora et al.\ reached the same conclusions as Larsen \& Richtler, i.e., NGC~4395 has a
small number of star clusters. Mora et al. estimated the ages and masses of the clusters detected,
showing that the star cluster system contains no objects with ages older than $10^9$ yr and includes
clusters with masses ranging from $200$ to $\sim3\times10^4$ M$_{\odot}$. 

Until now, no study of resolved field stars has been performed in this galaxy. 
However, \citet{LR00} found that NGC 4395 has one of the lowest area-normalised star formation rates
of all objects in their sample.


\section {Observation and data reduction}

Images of NGC 4395 were taken using the Wide Field Channel on the {\it Advanced Camera for
Surveys} (ACS/WFC) and the {\it Wide Field
Planetary Camera 2} (WFPC2), both on board the {\it Hubble Space Telescope} (HST). The resolution of the
detectors are $0\farcs05$, $0\farcs046$, and $0\farcs1$ per pixel for ACS/WFC, WFPC2/PC, and WFPC2/WF, 
respectively. At the distance of NGC 4395, $0\farcs05$ corresponds to a linear scale of $\sim 1$~pc.

Two different fields of the galaxy were observed, covering
the two spiral arms (see Fig. \ref{fig:ngc4395}). The
images were taken using the filters F336W ($\sim $ U) using WFPC2 and
F435W ($\sim $ B), F555W ($\sim $ V) and F814W ($\sim $ I) using ACS,
for each field. Each exposure was divided into two sub-exposures 
to eliminate cosmic-ray hits. For the ACS images, these sub-exposures were
also dithered to allow removal of cosmetic defects.
Table \ref{tab:data} summarizes the observations.

The data were processed with the standard STScI pipeline. The raw
ACS images were drizzled using the {\it multidrizzle} task \citep{koekemoer02}
in the STSDAS package of IRAF\footnote{IRAF is distributed
by the National Optical Astronomical Observatory (NOAO), which is
operated by the Association of Universities for Research in Astronomy,
Inc, under cooperative agreement with the National Science
Foundation}. The default parameters were used, but automatic
sky subtraction was disabled. The WFPC2 images were combined 
and corrected for cosmic rays using the {\it crrej} task with the default
parameters.

\subsection {Object detection}

At the distance of NGC~4395, the spatial resolution of the HST images allows
us to distinguish field stars (point sources with typical $FWHM\sim2$ pixels) and star clusters (extended sources
with typically $FWHM\sim3$ pixels or greater) in the images. We analyzed the two
separately and applied different detection criteria optimised for each type of
object:

\begin {itemize}
\item {\it Field star candidates (point sources)}\\ To detect
field stars, we created an averaged image (using the bands
B, V, and I) for each field. We ran the {\it daofind} task in IRAF
for the detection of stars using a $4\sigma$ detection threshold and a 
background standard deviation of $\sim0.02$ in units of counts per second.

\item {\it Star cluster candidates (extended sources)}\\ Cluster
detection was also performed on the averaged image.
The object detection was performed using SExtractor V2.5.0 \citep{bertinarnouts96}.
The parameters used as input for this program were 6
connected pixels, all of them with 10$\sigma$ over the background,
to remove point- and spurious sources as much as
possible. From the output file of SExtractor, we kept the coordinates
of the objects detected and the FWHM calculated by this program. The
coordinates found with SExtractor were then passed to {\it ishape} in
BAOLab \citep{larsen99}. {\it Ishape} models star clusters as analytical \citet{king62}
profiles (with concentration parameter $t_{tidal}/r_{core}=30$) and takes the instrument's PSF, created 
over the average image during the photometry, into account (see Sect. 4 for details on the creation of the PSF).
By minimization of a $\chi^2$-like function, {\it ishape} calculates the best-fit cluster coordinates, size, i.e., FWHM,
signal-to-noise ratio, and the $\chi^2$ of the best fit. From the output of {\it ishape} we saved the coordinate of the objects, 
the FWHM, the $\chi^2$ of the fit, and the signal-to-noise ratio (calculated within the fitting radius
of 4 pixels). 

For each WFPC2 chip, between five and seven common stars were visually selected
and used to convert the coordinate list from ACS to WFPC2 frames using the task {\it geomap} in 
IRAF. The transformations had an {\it rms} of $\sim 0.15$ pixels.

\end{itemize}


\section {Photometry}
The photometry of NGC 4395 was performed following standard aperture
and PSF fitting photometry procedures as described in the following.

\subsection {Field stars}
Because of the crowding in our fields, we performed PSF
fitting photometry to study the field stars. A set of bona fide stars
were selected by eye to construct the PSF (for each band a different
group of stars were used because the same stars might have different brightnesses in different bands). 
The PSF photometry was performed with DAOPHOT in IRAF.
We selected the PSF stars by measuring the FWHM (using {\it imexamine})
and selecting point sources smaller than $FWHM\approx2.2$ pixels. As
far as possible, we tried to include isolated stars distributed over the whole image.

The raw magnitudes were converted to the Vega magnitude system using the 
HST zero-points taken from HST webpages\footnote{http://www.stsci.edu/hst/acs/analysis/zeropoints/\#tablestart}
after applying aperture corrections to a nominal $0\farcs5$ aperture
(see sect. 4.4 for a more detailed description of how aperture corrections were determined).
The zero-points used were $ZP_B$ = 25.767, $ZP_V$ =25.727, and
$ZP_I$ = 25.520 magnitudes.

\begin{figure}[!t]
	\centering
	\includegraphics[trim= 0mm 0mm 0mm 10mm, width=\columnwidth,height=80mm]{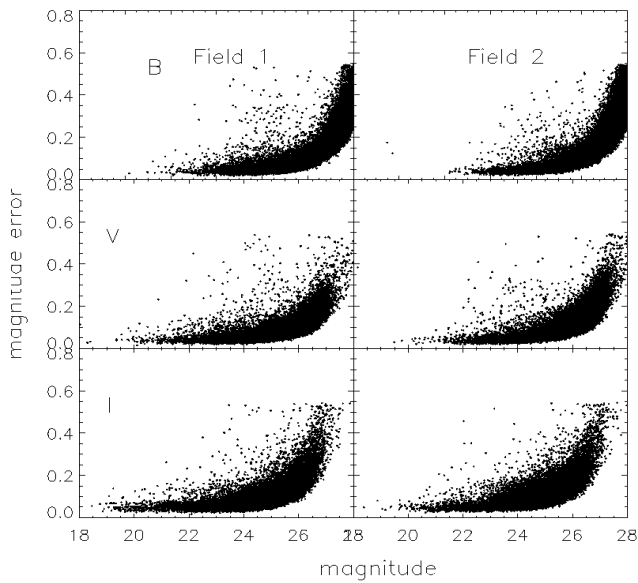}
	\caption{Magnitude errors for the stars detected using PSF
	         fitting photometry over the two fields for the bands
	         B, V, and I}
	\label{fig:magerrors}
\end{figure}

Figure \ref{fig:magerrors} shows the errors in our PSF photometry ($1\sigma$ error) versus
magnitude. The errors increase strongly below magnitudes of $\sim26$
in each band, corresponding to absolute limits of $\sim-2$ mag at the
distance of NGC~4395.

\begin{figure}[!t]
	\centering
		\includegraphics[trim= 16mm 10mm 0mm 5mm,width=\columnwidth]{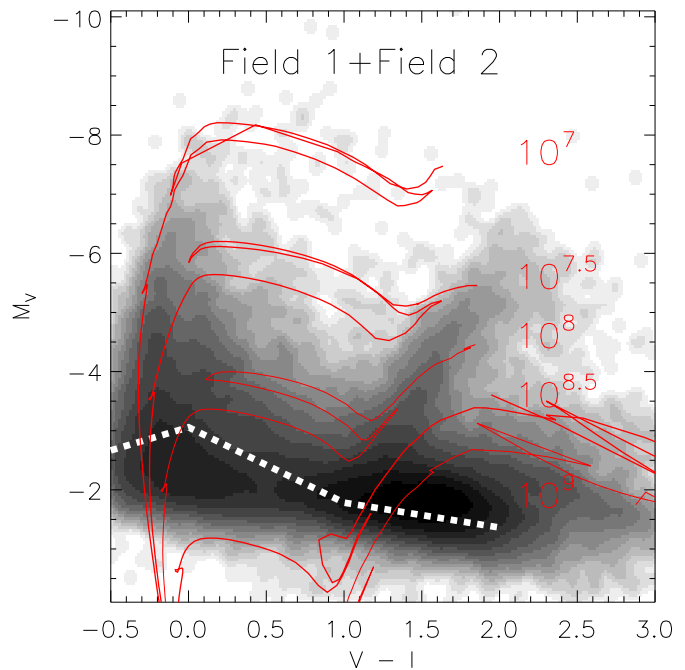}
		\caption{Hess diagram for the field stars in both fields. The dashed white line 
	         	represents the $50\%$ completeness  for the first field. Red lines are Padova 2008 theoretical isochrones
		for the ages $10^7$,$10^{7.5}$,$10^8$,$10^{8.5}$, and $10^9$ yr using LMC metallicity.}
	\label{fig:hess}
\end{figure}

A total of $\sim$30 000 stars were found in each field.  A combined Hess diagram for both fields 
is shown in Fig. \ref{fig:hess} (a Hess diagram plots the relative frequency of 
stars at different color-magnitude positions). Various phases of stellar evolution can be
recognised in the Hess diagram:

\begin{enumerate}
\item Main sequence and possible blue He-core burning  stars at $V-I \sim0$ and $-2\le V\le-8$;
\item Red He core burning stars at $1.2\le V-I \le2.5$ and $-2.5\le V\le-6.5$; 
\item RGB/AGB stars at $1\le V-I \le3$ and $-0.5\le V\le-2.5$.
\end{enumerate}

Overplotted in Fig. \ref{fig:hess} are theoretical isochrones from the Padova group \citep{marigo08}
for five different ages using metallicity $Z=0.008$ (red lines). The O abundance of
H{\sc ii} regions in NGC~4395 was measured by \citet{roy96} to be $12 + \log$O/H = $8.33\pm0.25$,
or about 1/3 solar \citep{grevessesauval98}, suggesting an overall metallicity
similar to that of the LMC.  This is consistent with our analysis in Sect. 5, which shows that isochrones
of LMC metallicity reproduce our data more closely. The white line indicates the $50\%$ completeness limit (see Sect. 4.3).

\subsection {Star clusters}
Our detection criteria is met by 16 463 objects, i.e., 6 connected pixels with 
10$\sigma$ above the local background level.
For the clusters, we carried out aperture photometry. An aperture
radius of 6 pixels was used for the ACS images. At the distance of NGC 4395, 1 ACS/WFC 
pixel corresponds to $\sim1$ pc, hence the chosen source aperture contains about 2 half-light 
radii for a typical star cluster. Our sky annulus had an inner radius of 8 pixels and a width of 5 pixels. 
For the WFPC2 images, apertures covering the same area were used (source aperture = 3 pixel radius, 
sky annulus = 4 pixel inner radius and 2.5 pixels width).

Of the 16 463 objects detected with SExtractor/{\it ishape}, a total of 4 472 candidates have measured four band
photometry.

ACS magnitudes were converted to vega magnitude system using the
same tables as for the field stars. WFPC2 magnitudes were converted to the Vega magnitude system using
the zero-points taken from the webpages\footnote{www.stsci.edu/instruments/wfpc2/Wfpc2\_ch52\#1933986}
of HST. Charge transfer efficiency (CTE) corrections were applied following
the equations from \citet{dolphin00}\footnote{Last update May 12,2008:\\ \url{http://purcell.as.arizona.edu/wfpc2\_calib/}}.

To select star cluster candidates, we used 3 criteria:

\begin{itemize}
\item Size:\\ 
We measured the sizes of the objects. Figure \ref{fig:fwhms}
shows the FWHM distributions for SExtractor and {\it ishape}.  The
SExtractor histogram peaks at $\sim 2.2$ pixels, 
corresponding to the stellar PSF. The {\it ishape} histogram peaks at 0 pixels, as {\it ishape}
takes the PSF directly into account. Based on the FWHM
distributions, we decided to use a criteria of $FWHM_{SExtractor} \geq
2.7$ pixels and $FWHM_{ishape} \geq 0.7$ pixels as a first selection 
of extended sources. These two limits correspond to a physical cluster
half-light radius of $\sim1$ pc or greater at the distance of NGC~4395.
Many of the candidate clusters we detect are low-mass and have irregular profiles often 
dominated by a few stars, so we chose to rely on both SExtractor and Ishape size measurements 
to achieve a more robust rejection of unresolved sources.

\begin{figure}[!t]
	\centering
		\includegraphics[width=\columnwidth]{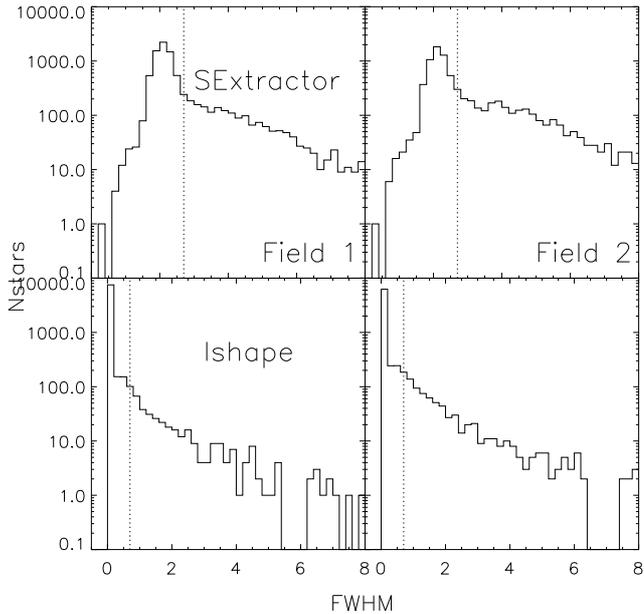}
		\caption{Histograms of the FWHM of the objects detected with
	         	SExtractor (top panel) and {\it ishape} (bottom panel). The
	         	vertical dotted lines represent the limits used to select the extended objects
	         	($FWHM_{SExtractor} \geq 2.7$ and $FWHM_{ishape} \geq
	         	0.7$ pixels)}
	\label{fig:fwhms}
\end{figure}

\item Magnitude:\\ 
\citet{mora09} ran a completeness analysis of artificial star
clusters of different FWHMs in five nearby galaxies, including
NGC 4395. The results presented in their work establish a $50\%$ limit
between $25\leq m_B\leq26$ magnitudes for objects with $FWHM=[0.1$(point sources)$,1.8]$
(see \citet{mora07,mora09} and Sect. 4.3 for more details).

By performing four band photometry, we  set a magnitude cut-off at
$m_V \leq 23$, which represents objects brighter than $M_V\sim-5 $ at the
distance of the galaxy. This limit is brighter than that
found by \citet{mora07} by $\sim2$ magnitudes. 

The main difference between the selection criteria of \cite{mora07} and this work is in the size limits 
of {\it ishape}, of half of a pixel (Mora et al. used $FWHM_{ishape} \geq 0.2$), hence we expect our data not
to be significantly affected by incompleteness to our magnitude cutoff. 
However, even at our limit of $M_V=-5$, there is a risk that a few stars may dominate
the light originating in a cluster, making it difficult to
differentiate a real cluster from a couple of stars that, by chance,
could be in the same line of sight.

\item Color:\\
Without taking into account any significant reddening, all clusters,
including globulars, will have colors bluer than $V-I\sim1.5$, e.g., 
\citet{forbes97}, \citet{larsen01}. We therefore make a color cut at $V-I=1.5$.

\end{itemize}

\begin{figure}[!t]
	\centering
		\includegraphics[width=\columnwidth]{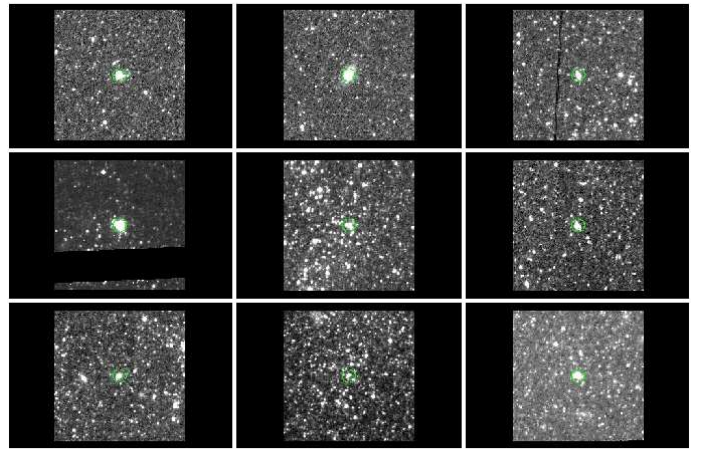}
		\caption{Example stamps of the clusters that satisfy the criteria stated in this work. Each cluster image has a dimension of $100\times100$
			      pixels ($\sim100\times100$ pc).}
	\label{fig:stamps}
\end{figure}

We wish to emphasise that there is probably no unique combination of criteria that will lead
to the detection of all bona-fide clusters in the image and at the same time produce no 
false detections. At low masses and young ages in particular, the light profiles may be dominated by individual bright
stars. Out of 4 472 candidate objects, a total of 22 objects fulfill the three 
criteria stated above and will be considerate in the remain of this study to be star clusters. 
These 22 clusters were visually inspected in our images to determine whether they resemble a cluster.
An example of the clusters detected is presented in Fig. \ref{fig:stamps}.
We investigated the large number of rejected objects and found that they were rejected for many reasons:
(1) the area covered by the two detectors differs by a factor of $\sim2$;
(2) many of the objects are too faint in the U band; (3) the magnitude cut-off removes many
objects, e.g., from a total of 4 475 detected objects, only 1 351 satisfy the magnitude limit, leading to the loss of $\sim70\%$ detections; and
(4) a large fraction of the objects have FWHMs smaller than our limits.

\begin{figure}[!t]
	\centering
		\includegraphics[width=\columnwidth]{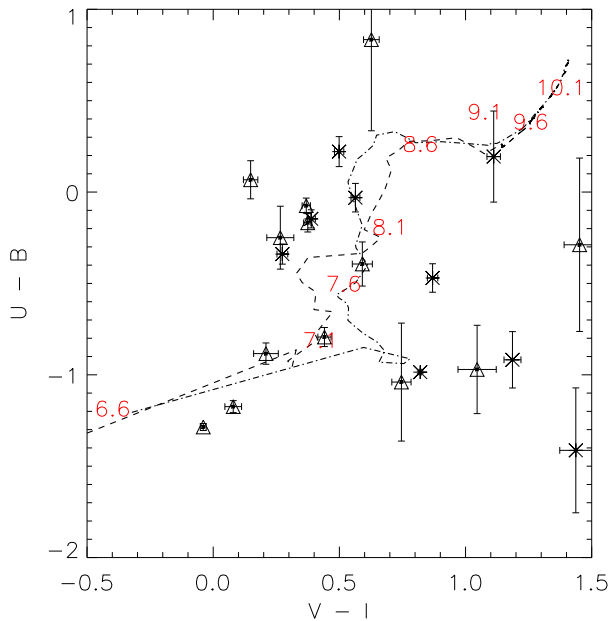}
		\caption{Two-color diagram for the star clusters detected in
	         	both fields, $ \times $ represents the first field and $ \triangle $ the second field. 
		The lines represent the theoretical GALEV
	         	track for a cluster with LMC (dashed line) or solar (dash-dotted line) metallicity, adopting Padova isochrones,
	         	a Salpeter IMF, and ages between $6.6\leq log(\tau/{\rm yr}) \leq10.1$}.
	\label{fig:twocolor}
\end{figure}

For the extended objects, a two-color diagram, based on the photometry 
in all of the available 4 passbands, is presented in Fig.
\ref{fig:twocolor}. Clusters in both fields are depicted with their respective
photometric errors and corrected for foreground extinction ($A_B=0.074$). Using GALEV models \citep{andersfritze03},
Padova isochrones \citep{bertelli94}, a Salpeter's IMF \citep{salpeter55}, and LMC 
(dashed line) or solar (dash-dotted line) metallicities, we overplotted the track that a cluster follows from
ages between $6.6\le Log(\tau)\le10.2$ yr (each $\Delta \log(\tau)=0.5$ dex age in log units being indicated). 
We observe considerable scatter in the two-color diagram. For these relatively low-mass
clusters, the discreteness of the initial mass function will be important and might contribute significantly to the scatter
\citep{girardi95,cervignoluridiana06,maizapellaniz09}. 
A more detailed study of stochastic effects will be presented in a forthcoming paper (Silva-Villa et al.\ 2010, in prep.)
This scatter was previously observed by \citet{mora09} (see their Fig. 6). Considering this
scatter, it is clear that ages derived by a comparing observed and model colors should
be treated with some caution.

\subsection {Completeness test}
\begin{figure}[!t]
	\centering
		\includegraphics[width=\columnwidth]{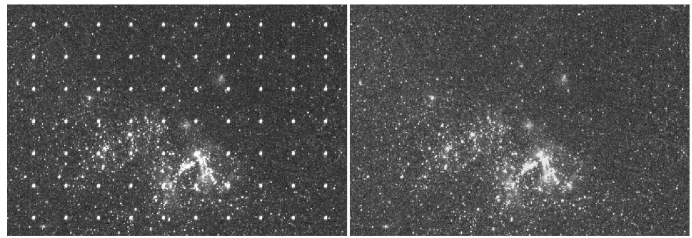}
	\caption{Subsection of the original image (right) and completeness image (left).}
	\label{fig:cmpl}
\end{figure}

To determine the completeness limits of our field-star photometry, we generated
synthetic images with artificial stars. From 20 to 28 magnitudes, in steps of 0.5 mag, 
five images per step were created and analyzed with exactly the same parameters used
in the original photometry. For each combination of color and magnitude,
 $528$ artificial stars were added using 4 different regions over the image to cover crowded and uncrowded areas.
These test images were created using the task
{\it mksynth} in BAOLab \citep{larsen99} and using the original PSF
images created during the photometry procedures (see Sect. 4.1). The 
separation between two consecutive stars was 100 pixels (without any sub-pixel variations), 
avoiding possible overlaping among the stars. The resulting test images were added to the science images using the task {\it imarith}
in IRAF. As an example, a subsection of an image, both test
and science, is presented in Fig. \ref{fig:cmpl}, where the fake stars have magnitudes
$m_V=21$.

\begin{figure}[!h]
	\centering
		\includegraphics[width=\columnwidth]{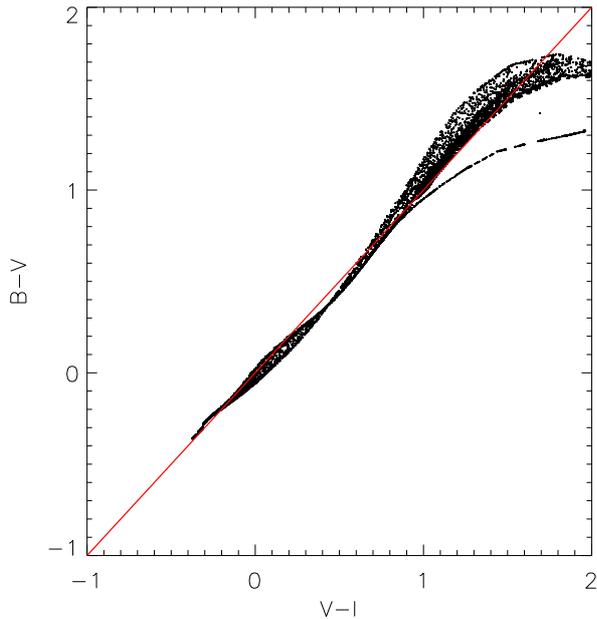}
	\caption{Two color diagram for theoretical values of B, V and I bands from Padova 2008 isochrones adopting LMC metallicity
		      for ages between $10^{6.6}$ to $10^{10}$ yr. The red line represents a $1:1$ relation between the colors B-V and V-I.}
	\label{fig:11relcmpl}
\end{figure}

To performe a realistic completeness analysis it is in principle necessary to sample the three-dimensional 
($B,V,I$) color space. However, since different colors are tightly correlated with each other, the problem
can be reduced to a two-dimensional one. Figure \ref{fig:11relcmpl} was generated using Padova 2008 
isochrones \citep{marigo08}, assuming solar metallicity, in the color range from -1 to 2, for $B-V$ and $V-I$
and shows that there is a nearly 1:1 relation between these two colors (the red line in Fig. \ref{fig:11relcmpl} represents a 1:1 relation, 
but not an accurate tof the data). We created the images for the completeness test described above using this approximation between the three bands, i.e., 
if a $B$ image has stars with $m_B=21$ and $B-V=1$, then the $V$ and $I$ images will have stars
of $m_V=20$ and $m_I=19$, respectively, allowing us to perform a study of magnitudes and color variations
over the images.

\begin{figure}[!h]
	\centering
		\includegraphics[width=\columnwidth]{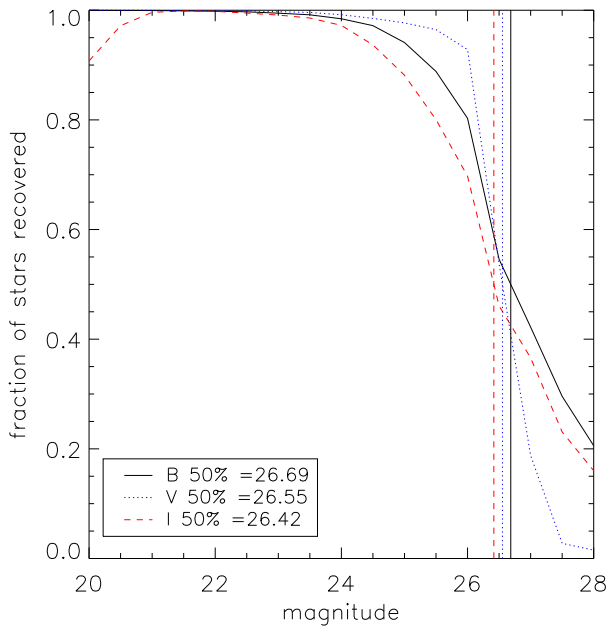}
	\caption{Completeness diagrams for the bands B,V, and I studied over the first field. Vertical lines represent the $50\%$ completeness.}
	\label{fig:cmpl-pos1}
\end{figure}

Based on these tests, we found an average of $50\%$ completeness limits for the whole color range at
$m_B = 26.69$, $m_V = 26.55$, and $m_I = 26.42$ for the first field and $m_B =
26.71$, $m_V = 26.49$, and $m_I = 26.39$ for the second field. Figure
\ref{fig:cmpl-pos1} shows the completeness diagram obtained from
this analysis for the first field (for the second field, the figure is
similar) and each magnitude.

The color-dependent 50\% completeness limit is shown as a white dashed line
in the Hess diagram (see Fig. \ref{fig:hess}). Since the completeness functions are
very similar for the two fields, we can combine the photometry for both fields and
use just one set of completeness tests in the following analysis.

We did not performe a completeness test for star clusters but refer to the tests
performed by \citet{mora07,mora09}, who used the same data and very similar
cluster detection procedures. These authors created artificial star clusters using 
different FWHMs from 0.1pixels (stars) to 1.8 pixels and a range of magnitudes from 16 to 26 for three square grids in different
positions over the image, trying to cover crowded and non-crowded areas. 
To create the fake clusters, they assumed a \citet{king62} profile
with $r_{tidal}/r_{core}=30$. Using the {\it mkcmppsf} task on BAOLab, 
fake extended objects were added to an empty image and then added
to the science image to performe measurements. The selection criteria
in \citet{mora07,mora09} is a $FWHM\ge(2.7,0.2)$ pixels for SExtractor and {\it ishape}, respectively.
For high background levels, Mora at al. found a shallower detection limit; nevertheless, all
the limits correspond a 50\% completeness limit between $m_B\approx25$ and $m_B\approx26$,
2-3 magnitudes fainter than the cut at V=23 that we apply for the selection of cluster candidates.

\subsection{Aperture corrections}
The aperture corrections were determined separately for field stars
and star clusters. The aperture corrections for the field stars were derived
following standard procedures, while for extended objects we
adopted the relations found by \cite{mora09}.

\begin{itemize}

\item Field stars:\\ 
By ``aperture corrections'', we here mean corrections from the PSF-fitted
instrumental magnitudes to aperture photometry for nominal radii of $0\farcs5$.
These were measured using a set of isolated visually selected stars across the 
images. From $0\farcs5$ to infinity, we applied the \citet{sirianni05} values. The
corrections obtained are of the order $\sim 0.1$ mag (see table 
\ref{tab:aperturecorrections}).

\item Star clusters:\\ 
\citet{mora09} estimated a relation between the aperture corrections and the sizes (FWHM) of star clusters
using the same data set used in this paper. Photometric parameters in both \citet[][and ours]{mora09} work
are the same, allowing us to assume the relations found in their work 
and apply these aperture corrections (size-dependent) to our data, following eq. 1 in \citet{mora09}.
This set of equations is also band-dependent, although we used the sizes of the objects
measured on an average image. 

The aperture corrections by \citet{mora09} correspond to a nominal aperture of $1\farcs45$. From this nominal aperture
to infinity, we adopted the values presented by \citet{sirianni05}, although these corrections are $\sim0.03$ 
magnitudes for the bands B, V, and I (within $1\farcs5$ about 97\% of the total energy is encircled).

\end{itemize}

\begin{table}[!t]
	\centering
	\caption{For point source. B, V, and I aperture corrections  to a nominal aperture of $0."5$, estimated in this study.}
		\begin{tabular} {c c c c} 
			\hline \hline
			 & $B_{F435W}$ & $V_{F555W}$ & $I_{F814W}$ \\ 
			 & [mag] & [mag] & [mag] \\ \hline \hline
			{\it Field 1} & 0.07 & 0.03 & 0.08 \\
			{\it Field 2} & 0.05 & 0.08 & 0.06 \\
			 \hline \hline
		\end{tabular}
	\label{tab:aperturecorrections}
\end{table}


\section{Star formation histories, ages, and masses}

Our main goal in this paper is to compare the field stars with the star
cluster populations in NGC~4395. In the following section, we describe how
we derive the star formation histories (SFHs) of the field stars and the
ages and masses of the clusters. We then proceed to
compare the two.

\begin{figure*}
	\centering
		\includegraphics[width=150mm]{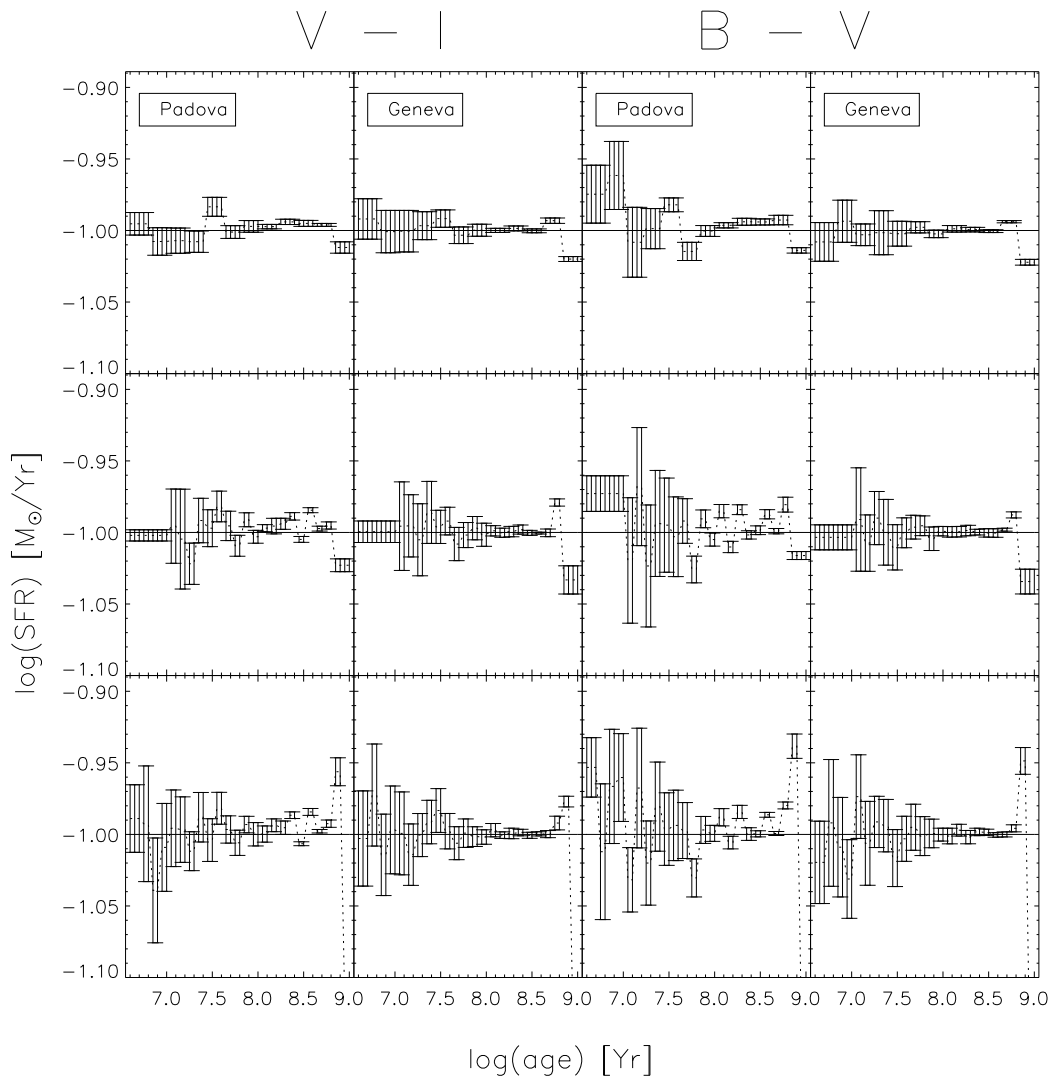}
		\caption{SFH test using the bands B, V, and I. Artificial populations were created using Padova and Geneva isochrones
		and then passed to the program. The procedure was performed 10 times and the mean SFH (dashed line)
		and the respective standard deviation is shown, relative to the input assumed SFH of $0.1$ M$_{sun}$yr$^{-1}$
		(straight line). Each row represents a different binning used for the study, which decreases in size from top to bottom.}
	\label{fig:fakeall}
\end{figure*}

\subsection{Deriving star formation histories: approach and its testing}
To estimate the SFH of the field stars, we have used the synthetic CMD method. 
The synthetic CMD method consists of creating
of an artificial photometric distribution of stars taking into account
photometric errors, distance modulus, IMF,  binarity, and
interstellar extinction.  The relative weights of the stellar isochrones
used in generating the synthetic CMD are adjusted
until the closest possible match of the synthetic CMD to the observed
one is obtained. These weights can then be translated to a
SFH.

Following the work of \citet{tosi91}, \citet{dolphin97}, and \citet{aparicio97},
many authors have implemented this technique. Some examples are 
STARFISH by \citet{harriszaritsky02}, IAC-Star
by \citet{apariciogallart04}, IAC-pop by \citet{apariciohidalgo09}, among others.
\citet{coimbra02} presented the results of a workshop (The Coimbra Experiment)
devoted to comparing different methods and interpretations of the
results from various groups using a homogeneous data set and physical inputs. The results
showed a large scatter at young ages, but good agreement at old ages.

We implemented the synthetic CMD technique ourselves as
an IDL program. We chose to develop our own code to include any 
features that we might wish (or need), and there is no clear preference in the literature 
for any of the other codes. The IDL code used in this work was based on our previous experience 
with the synthetic CMD technique \citep{larsen02}.
The program searches for the best-fit synthetic Hess
diagram using a maximum-likelihood technique. We assume that
the likelihood of observing $k_i$ stars in the $i$th pixel of the Hess diagram is given by
the Poisson distribution $\mathcal{P}(k_i; \mu_i)$, where $\mu_i$ is the number of stars expected in this pixel according to a given 
model Hess diagram. The program then searches for the linear combination
of input isochrones that maximises the total likelihood $\mathcal{L}$ over
the entire Hess diagram
\begin{equation}
  \log \mathcal{L} =  \sum_i \log \mathcal{P}(k_i; \mu_i).
\end{equation}
In practice, we assign a low (non-zero) probability of having a star even
in pixels where $\mu_i = 0$. This is essentially equivalent to allowing a
constant ``background level'' of stars that are not fitted by any isochrone. 
This is necessary since any star in a pixel with $\mu_i=0$ would otherwise
immediately drive the total likelihood to zero. Since no combination of 
existing isochrones provides a perfect match to the observed CMDs, and one
may in any case always expect some contamination of the CMD by foreground
stars, background galaxies, etc., this would be overly restrictive.
We also experimented with other criteria, e.g.\ minimum r.m.s. difference between
observed and model Hess diagrams, but tests similar to those described in Sect. 5.2 
below indicated that the maximum-likelihood method gave the most reliable results.


\begin{figure}[!t]
	\centering
		\includegraphics[width=\columnwidth]{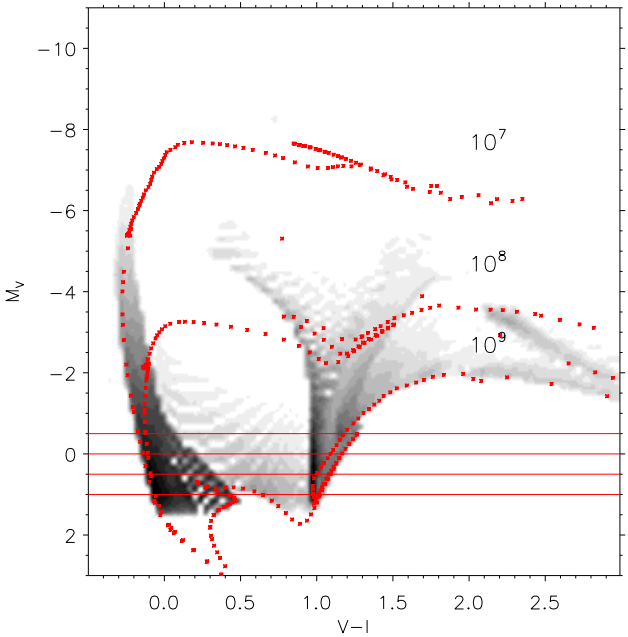}
		\caption{Hess diagram for the 4 possible completeness limit studied in Fig. \ref{fig:cmpls}.}
	\label{fig:hesscmpls}
\end{figure}

To create the synthetic CMD, the program uses theoretical isochrones either
from Padova \citep{marigo08} or Geneva \citep{lejueneschaerer01} or any other 
set of isochrones, as long as they tabulate the relevant color versus mass. 
If the weight of each isochrone were fitted independently, this would lead to a very large
number of free parameters, due to the small difference in age between
individual isochrones, e.g. $\Delta \log\tau=0.05$ for the Padova isochrones. 
We therefore group the isochrones together in age bins that can be defined
by the user, typically using a bin size of $\Delta \log \tau = 0.15$ dex.

The program allows us to assign different weights to different rectangular regions of the CMD,
selected by the user. This is useful because some phases of stellar evolution are more uncertain than
others, e.g. the blue loop stars, and thus should carry less weight in the overall 
determination of the SFH. The use of completeness limits (if known) can be applied before
the creation of the synthetic models, interpolating the completeness limits over
the whole area of the CMD. The option of applying the completeness correction after the creation of the synthetic model is simply 
a multiplication of each pixel in the synthetic CMD with a number between 0 and 1. We also include a simplified 
treatment of binaries, in which binary evolution is ignored but the effect of unresolved binaries on the CMD are modeled.
The program currently allows three different assumptions about the mass ratios in binary
systems, namely a delta function, an IMF sampling and a flat distribution. The metallicity, (a range of) extinction 
values, and distance modulus must also be specified. We allow the extinction to be age-dependent, 
by providing a list of ages and a range of extinction values for each age. The program then interpolates 
in this list for each isochrone.

\begin{figure}[!t]
	\centering
		\includegraphics[trim= 0mm 0mm 10mm 6mm,width=\columnwidth]{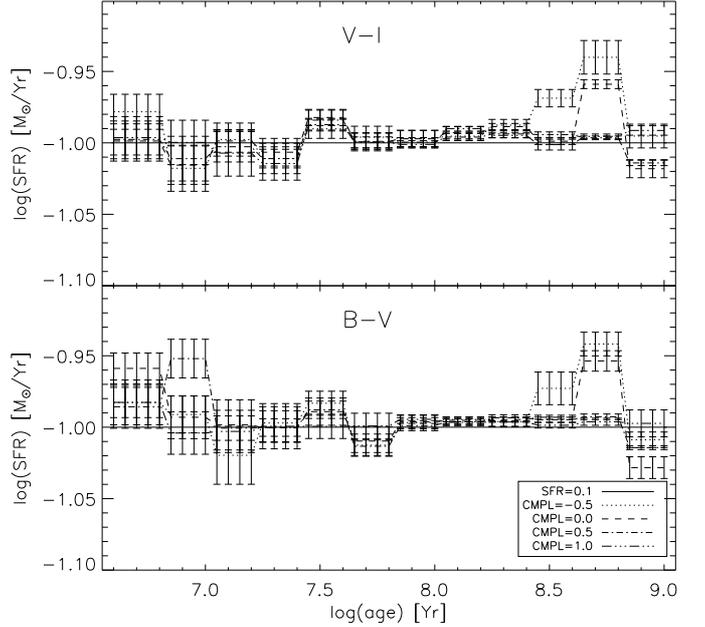}
		\caption{Test of the SFH using different completeness limits.}
	\label{fig:cmpls}
\end{figure}

\begin{figure*}
	\centering
		\includegraphics[width=150mm]{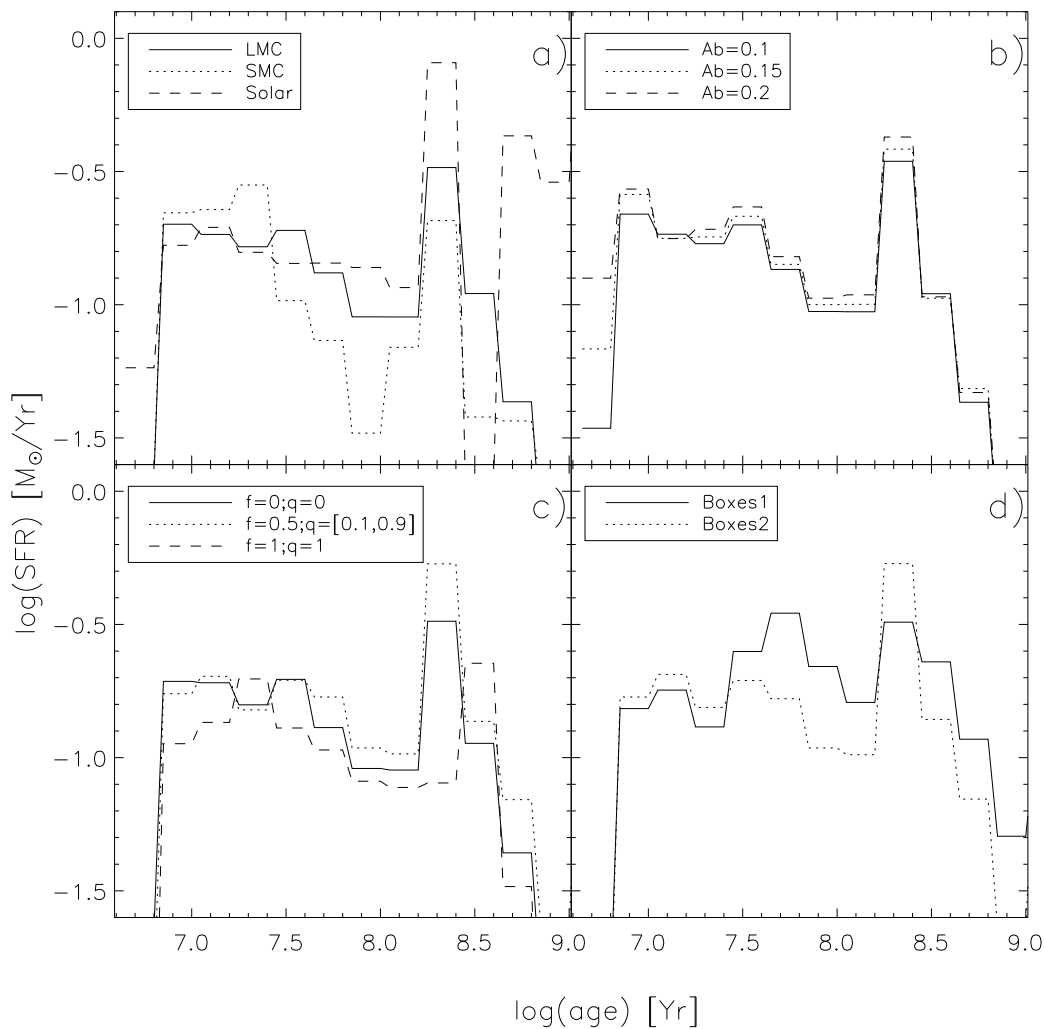}
		\caption{Padova theoretical isochrones were used to test the SFH program developed using the 
		photometric data obtained in Sect. 4. Panel {\it a)}: Test for different 
		metallicities (solar, LMC, SMC) without binarity or extinction assumed. Panel {\it b)}: Test for different extinctions 
		assuming LMC metallicity and no binarity. Panel {\it c)}: Test for different binarity assuming LMC metallicity and no extinction.
		Panel {\it d)}: Test for different boxes used over the fitting data assuming LMC metallicity, no extinction, and no binarity.}
	\label{fig:realtest}
\end{figure*}

Isochrones are populated by assuming a \citet{salpeter55} IMF with SFHs normalised to a mass
range specify by the user ($M=[0.15,100]$ M$_\odot$ is the default range).  Since we use Hess diagrams, each point in the CMD is a 
density function. To construct them, different kernels, namely square, Gaussian, disc, or delta function,
can be used. The kernels are of adjustable resolution and dimension. 
In the test presented below, we adopted a resolution for the Hess diagrams of
$100\times100$ pixels and used a delta function kernel. Each isochrone is broadened by 
the assumed photometric errors and binarity, and shifted and broadened by the (range of) specified extinction values.
To ensure a smooth coverage of the CMD, the program interpolates the isochrones by
a factor of 10. After creating of the Hess diagram for each individual isochrone, the program
linearly combines them and a synthetic CMD Hess diagram is obtained.


We performed a number of tests to check the internal consistency of the program,
as well as the sensitivity of the derived SFHs to the various parameters involved.

\subsection{Internal consistency check of the code}

We created artificial stellar populations, passed them to the program,
and then checked how the output compared to the input. The populations were constructed
assuming a constant SFR of $0.1$ M$_{\odot}$yr$^{-1}$, Padova or Geneva theoretical isochrones,
a Salpeter's IMF, and solar metallicity. The mass of each star was randomly chosen using Salpeter's
prescription for masses between 1 and 100 M$_\odot$ (low-mass stars ($\leq 1$ M$_\odot$) are too faint to appear in the Hess diagram,
especially after applying a completeness limit). Because we restricted the mass range
in our analysis, we then extrapolated to the default mass range used by the code ($M=[0.15,100]$ M$_\odot$).
The age of each star was assigned randomly (from a uniform distribution) for ages between 4 Myr to 1 Gyr.
Having the mass and the age of each star, magnitudes were obtained by interpolating the
theoretical isochrones. The magnitude limit used in these tests was $M_V = 0$.
We did not include binaries or extinction in this initial internal consistency check of the code. 

Figure \ref{fig:fakeall} presents the results of this analysis. In all the panels, the input SFH is shown 
as a horizontal straight line and the average reconstructed SFH of the output after 10 different 
runs as dashed lines. Each run had a new population, i.e. different random realization, for the bands B, V, and I.
The error bars illustrate the standard deviation in the reconstructed SFHs. We studied the variations
in these results based on the different assumptions of binning. Three different bins were used in this test corresponding
to each row of Fig. \ref{fig:fakeall}.

\begin{figure}[!t]
	\centering
		\includegraphics[width=\columnwidth]{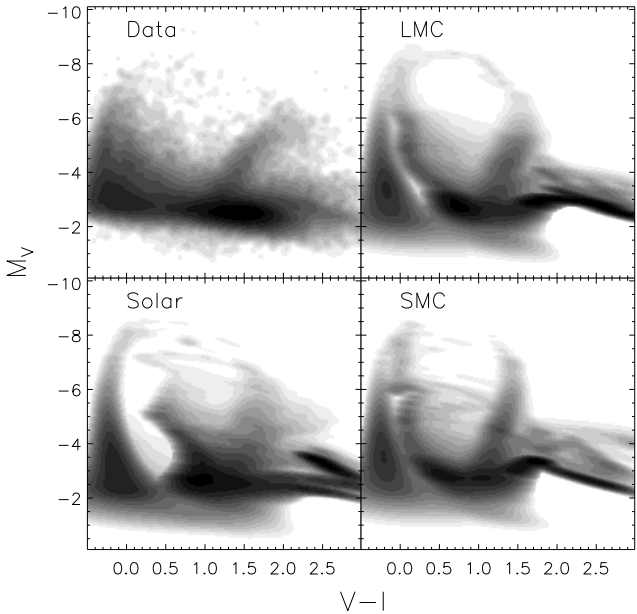}
		\caption{Comparison of the fit for different metallicities (solar, LMC, and SMC).}
	\label{fig:lmcsolar}
\end{figure}

We see that the reconstructed star formation histories generally agree fairly
well with the input values regardless of the binning, color, or isochrones used. It is observed
that the error bars are larger for smaller bins, although good agreement with the input SFH is maintained.
The maximum difference is $\Delta \log(SFR) \sim 0.05$ dex,  and generally far less.

We performed a test to see how completeness affects our SFH estimation.
Figure \ref{fig:hesscmpls} shows the Hess diagram of a stellar population with ages
between $10^6$ to $10^{9}$ yr using the mass range from above. Overplotted are 3 Padova 2008 isochrones
for ages $10^7,10^8$, and $10^9$ yr and four completeness limits  at $M_V = -0.5,0.0,0.5$, and $1.0$ 
magnitudes. Figure \ref{fig:cmpls} shows that the SFHs can be recovered reliably to
progressively older ages as the completeness limit becomes fainter. In particular, the ``burst'' at
$10^{8.8}$ years disappears when fainter stars are included in the fit.

\citet{cignonitosi09} created artificial CMDs to simulate a stellar population with a constant SFR between $0-13$ Gyr and used four
different completeness limits. They concluded that to safely reconstruct a SFH over a full
Hubble time from a CMD, we need to resolve all stars down to $M_V=4.5$. However, this completeness 
limit can only be reached for galaxies in the Local Group.

We conclude that our program produces consistent results at the level of 0.05 dex,
the age range over which this holds being restricted mainly by the completeness limits. This
\emph{internal} consistency of the code does not, however, imply that real SFHs can be
recovered with similar accuracy.

\subsection{Tests of sensitivity to assumptions about parameters}

We now concentrate on testing the effects of four parameters:
metallicity, extinction, binarity, and the fitting area used (rectangular boxes). To perform this test, we executed the code 
using the scientific data described in Sect. 4, using different assumptions
about these parameters and a mass range $M=[0.10,100]$ M$_\odot$ (which is the same mass range
used in Sect. 5.4). The results are shown in
Fig. \ref{fig:realtest}. Each panel represents one of the tested variables.

\begin{figure}[!t]
	\centering
		\includegraphics[width=\columnwidth]{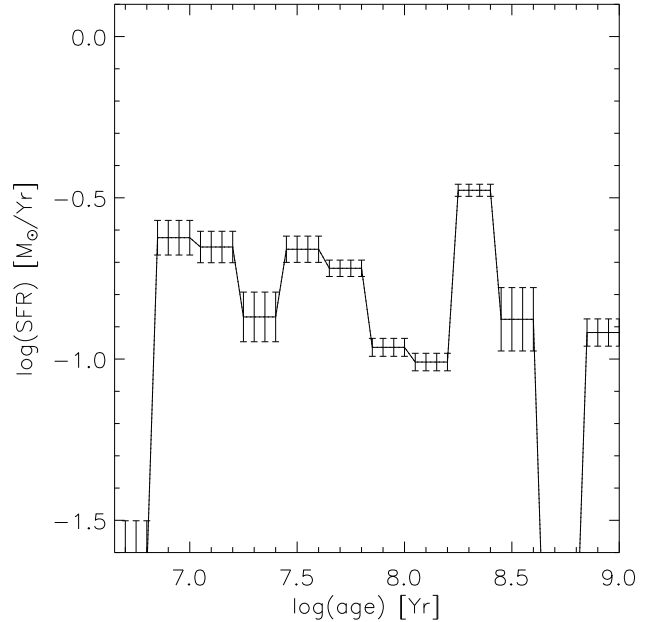}
		\caption{Bootstrapping study. The errors retrieved from this test are the random errors
			      of our program. No strong variations caused by random errors affect our results.}
	\label{fig:boostraping}
\end{figure}

In panel {\it a)}, we show the SFHs derived for three assumptions about
metallicity: solar (Z=0.02), LMC (Z=0.008), and SMC (Z=0.004). In this panel, we 
ignore binaries, include no additional extinction and used the fitting area called "boxes2" described below. We observed that the SFH
does not change very much from LMC to solar, but at SMC
metallicity there is a much stronger increase from $\sim100$ Myr ago to the present. By comparing the observed
and model Hess diagrams, it is clear that the solar metallicity isochrones
generally provide a poor fit, especially for the red and blue supergiants that
appear much too cool compared to the observations, while LMC and SMC metallicities are
very similar (see Fig. \ref{fig:lmcsolar}). 

Panel {\it b)} presents the behavior of the SFHs for three different assumptions about the total
extinction (foreground plus internal): $A_b=[0.1,0.15,0.2]$. We fixed the metallicity at $Z=0.008$ 
(LMC), again ignoring binaries, used Padova isochrones and used the fitting area "boxes2".
In general, the SFHs obtained using different extinctions 
do not vary strongly for these relatively modest variations in the extinction.
Greater extinction variations would produce larger shifts towards the red in the 
model Hess diagrams and are inconsistent with the data.

\begin{figure}[!b]
	\centering
		\includegraphics[width=\columnwidth]{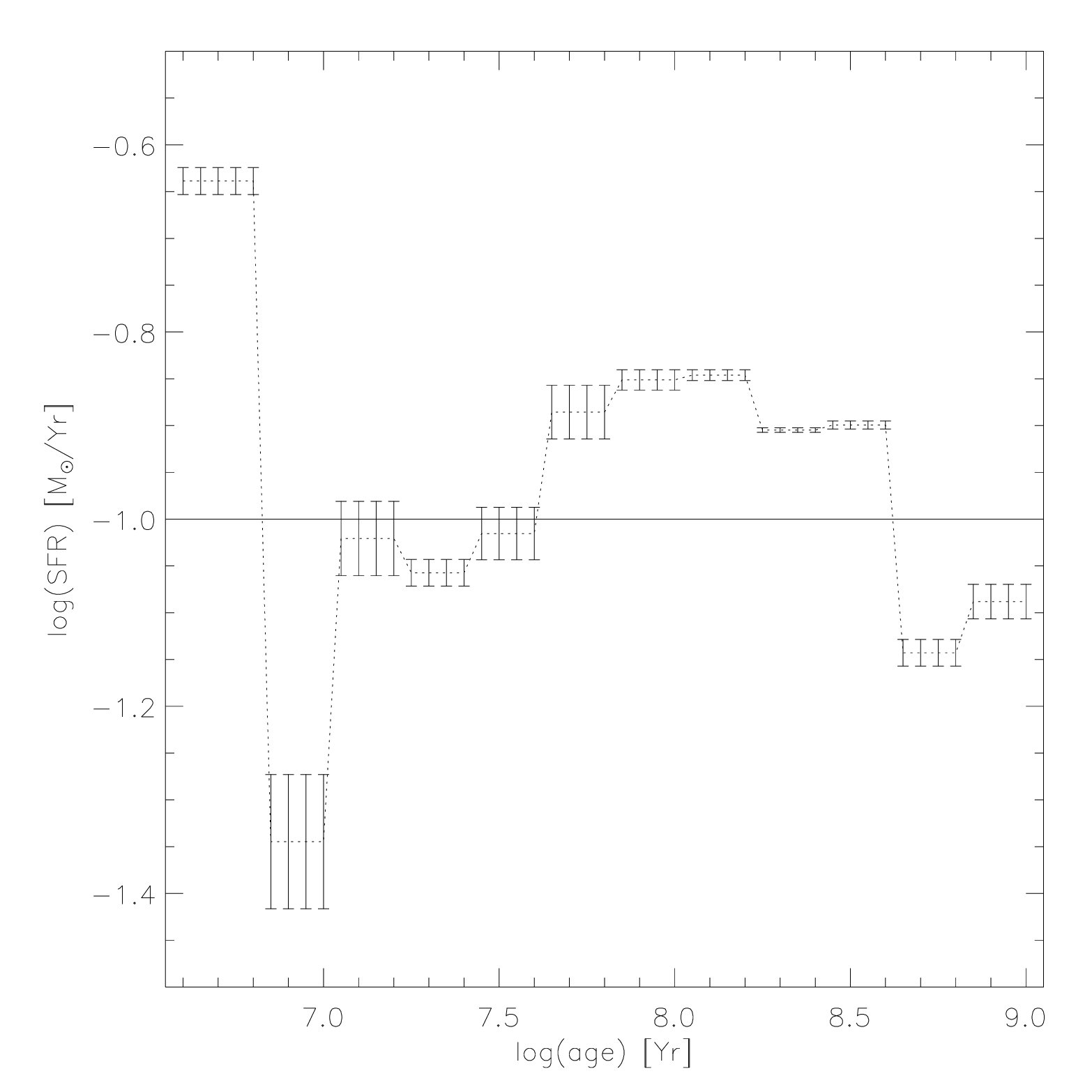}
		\caption{Test using Padova isochrones to create the population and Geneva isochrones to analyze them.}
	\label{fig:pad_gen}
\end{figure}

Panel {\it c)} shows the dependence on the assumptions for binary fraction ({\it f}) and mass 
ratio ({\it q}) of values (1.) $f=0$, $q=0$;  (2.) $f=0.5$, $q=[0.1,0.9]$; (3.) $f=1$, $q=1$. The panel also indicates how this dependence affects 
the estimation of the SFHs. We used Padova isochrones, fixed the metallicity  at $Z=0.008$, ignored
the effects of additional extinction, and used the fitting area named "boxes2" described below. Binarity can have an effect on the derived
SFHs with minor effects for $\tau\leq100$ Myr and a shift towards somewhat lower overall SFR for the extreme case $f$ and $q$ equal 1.
However, the binarity assumptions made here are not completely
realistic. The first and third assumptions are extreme cases with no binaries at all and a galaxy where
all the stars have a companion of exactly the same mass, respectively. 
The second assumption is an intermediate case where there is a continuous (flat) range of
$q$ values and a star has $\sim50\%$ probability of being in a binary system. However, in a more
realistic case binary \emph{evolution} should be taken into account as well. This is however beyond
the scope of the present work.

\begin{figure}[!t]
	\centering
		\includegraphics[trim= 0mm 0mm 0mm 0mm,width=\columnwidth]{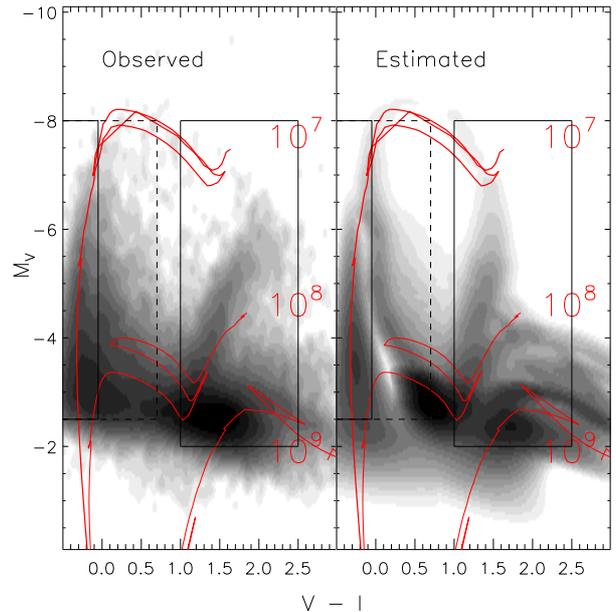}
		\caption{Comparison between the observed and fitted Hess diagrams for the field stars studied in this work.
		      As age indicators, three isochrones are overplotted at the ages of $10^{7}, \ 10^{8}$, and $10^{9}$ yr.
		      Boxes used to fit the data are represented for both Hess diagrams, observed and fitted.}
	\label{fig:hesscompare}
\end{figure}

In the last panel, panel {\it d)}, we show the SFH estimates obtained using two different sets of boxes in the fitting.
As overplotted in Fig. \ref{fig:hesscompare}, where we plot the observed Hess diagram and the regions fitted, 
the left box ($-0.5 \leq V-I \leq -0.05$ and $-8.0 \leq V \leq -2.5$) used contains the main sequence stars, the blue He core burning phases being outside of the fit, 
while the right box ($1.0 \leq V-I \leq 2.5$ and $-8.0 \leq V \leq -2.0$) contains the red He core burning 
stars and possible RGBs and AGBs. This set of boxes is called "boxes2". The second set of boxes covers the same area for the evolved stars (right box), while
the blue He core burning stars are also included in the fitting box ($-0.5 \leq V-I \leq 0.7$), which we label "boxes1" in 
panel {\it d)}. In general, we found that,  the SFH result from "boxes2" shows a lower value that for "boxes1". We
studied these two possible sets of boxes because the fitted Hess diagrams retrieved using these two boxes
indicated that the fit for the blue He core burning stars is not in good agreement with observations. 
The variations after $\sim\tau\sim10^{8.7}$~yr are unlikely to be real, but are probably due to
incompleteness as discussed above.

We created a fake population using Padova isochrones and assuming solar metallicity, no
extinction, and no binarity; we then analyze this with our program, using Geneva isochrones 
to see whether the assumed SFH could be recovered. Figure \ref{fig:pad_gen} shows the result 
of this test. There are significant differences between the input and recovered SFH, deviations 
being as large as $\sim 0.4$ dex at young ages ($<10$ Myr) and differences being at the level of 0.1-0.2 dex at older ages.
The error bars presented in Fig. \ref{fig:pad_gen} were determined using the same method used in the previous 
section, i.e., are equivalent to the standard deviation of the reconstructed SFHs after 10 different runs with different random realizations.

We finally test how random errors (due to the finite number of stars in the Hess diagram)
affect the SFHs, using a bootstrapping method whereby the input data, i.e., the photometric data found in this work, are randomly
resampled many times. To perform this test, we used the observations obtained in this work (see Sect. 4), 
Padova 2008 isochrones, LMC metallicity, no additional extinction, and no binarity.
Disabling binarity will not influence our results dramatically
and will save us computational time. We ran the program 100 times, redistributing randomly 
the photometric data in each run. From Fig. \ref{fig:boostraping}, we conclude that Poisson variations
do not significantly affect our results.

\begin{figure}[!t]
	\centering
		\includegraphics[trim= 0mm 0mm 0mm 0mm,width=\columnwidth]{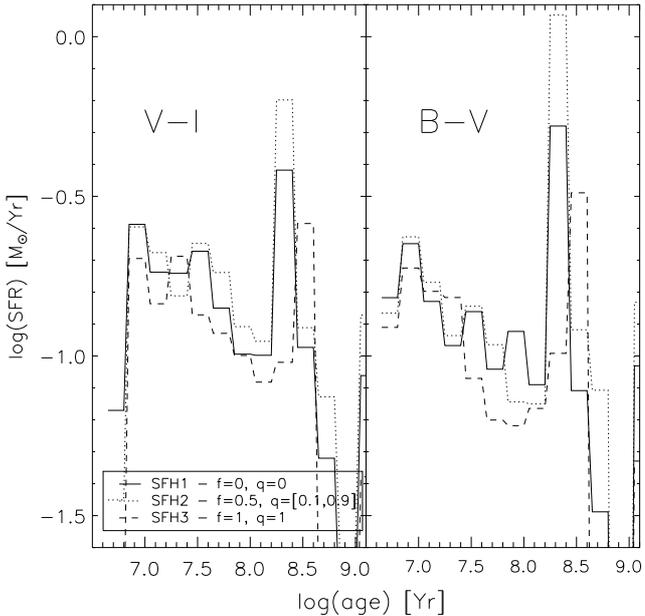}
	\caption{SFH for the galaxy NGC 4395. We studied the SFH for three different assumptions about the binaries trying
	to cover possible combinations between binary fraction and mass ratio for the colors V-I(left panel) and B-V(right panel).}
	\label{fig:sfhngc4395}
\end{figure}

To summarize, we demonstrated that the main uncertainties in our derived SFHs are systematic,
depending on assumptions about binarity, the choice of isochrones, and metallicity. However, 
with the exception of the use of solar metallicity isochrones (which provide a poor fit to the
data), the overall mean SFRs are fairly consistent with our data up to ages of a few hundred Myr.

\subsection{Results}

To infer the SFH of NGC~4395, we combined both fields to be able 
to directly compare with the cluster age distribution. The differences in 
our completeness limits for each band and each field are smaller than 0.1 magnitudes,
allowing us to use the completeness values obtained in our first field (results for 
the second field can also be used instead) to study the SFH of the whole galaxy. 
In the SFH reconstruction, we used a resolution of $200\times200$ pixels for
the Hess diagrams and a Gaussian kernel with a standard deviation of 0.02 mag in $V-I$
(or $B-V$). Based on our observed Hess diagram we used 2 rectangular fitting areas ({\it boxes2}) for the SFH reconstruction
with the same limits as described in the previous section. 
To check the consistency of our results we performed the analysis using both
$B-V$  and $V-I$ colors, using the same parameters and boxes to fit the data.

Figure \ref{fig:hesscompare} depicts the observed (left) and best-fit model (right) Hess diagrams (assuming $f=0.5$ and $q=[0.1,0.9]$).
The final combination of parameters used to estimate the SFH
were Padova 2008 isochrones, a Salpeter IMF (M=[0.10,100] M$_\odot$), LMC metallicity, three different assumptions 
for the binarity, distance modulus of 28.1, a total extinction of $A_B=0.15$, and the photometric errors
and completeness limit obtained in Sect. 4.

The phases that are most accurately reproduced by the fit are the main sequence, red core-He burning,
and RGB/AGB phases. For the blue core-He burning 
phases, the model fit is less good, being redder than observed and with a more pronounced
gap between the main sequence and blue core-He burning stars in the model than in the data.
This is a generic problem with the isochrones that exists for any star formation history we adopt.

Our most accurate estimates of the extinction were obtained by trial and error. We compared the main sequence stars from
the fitted Hess diagram with data for those observed and assumed different values ($A_B=[0.1,0.15,0.2]$) for the total extinction
until we obtained the best fit between the two Hess diagrams (considering the main sequence only).
The difference between the foreground value ($A_B = 0.074$) and our best-fit model value ($A_B \sim 0.15$)
suggests a low internal extinction in NGC~4395 of  $A_B\approx0.076$ mag. This estimation
was performed for LMC metallicity.

The estimated SFHs (for $V-I$ and $B-V$) are shown in Fig. \ref{fig:sfhngc4395}. 
As discussed in the previous section, the SFHs become very uncertain at ages greater than a
few hundred Myr; the apparent burst at $\sim10^{8.4}$ yr is probably not real. There is also a hint
of a rapid drop at very young ages in V-I ($\sim10^{6.8}$ years), but this may be caused by
uncertainties in the isochrones. Furthermore, the youngest age included in our artificial
Hess diagrams is 4 Myr; if younger stars were present in the field, these would be included
in the youngest bin. The appropriate lower age limit depends on how long young stars
are embedded in their native molecular clouds. We present the SFHs for different binarity assumptions
because this parameter is also uncertain. We use the same 3 choices in Fig. \ref{fig:sfhngc4395} as
those used in Fig. \ref{fig:realtest} above. Taking the values for each bin (in linear units), we estimated 
the average star formation rate for ages between $10^7$ and $10^8$ years 
using the colors V-I and B-V, as shown in Table \ref{tab:sfhsbincol}. 

\begin{table}[!t]
	\centering
	\caption{Average star formation rate [in M$_{\odot}$Yr$^{-1}$] over the past $10^8$ years, estimated using different binarity, colors,
			and isochrones.}
		\begin{tabular} {c c c c} 
			\hline \hline
			 & $f=0$ & $f=0.5$ & $f=1$ \\ \hline
			 & $q=0$ & $q=[0.1,0.9]$ & $q=1$ \\ \hline
			 & & Padova & \\ \hline
			V-I & 0.18 & 0.19 & 0.15 \\ 
			B-V & 0.14 & 0.14 & 0.12 \\ \hline \hline
			 & & Geneva & \\ \hline
			V-I & 0.20 & 0.21 & 0.19 \\ 
			B-V & 0.16 & 0.20 & 0.16 \\ \hline \hline
		\end{tabular}
	\label{tab:sfhsbincol}
\end{table}

There is an apparent increase in the SFR between $\log \tau \sim 10^{8.0}$ and $\log \tau \sim 10^{7.0}$, 
of $\sim0.5$ dex. Based on Fig.~\ref{fig:sfhngc4395},
the average star formation rate over the past $10^8$ years is $SFR\approx0.17$ M$_\odot$yr$^{-1}$
within the two ACS fields (assuming $f=0.5$ and $q=[0.1,0.9]$). We note that this is a factor of 4 higher than the \emph{global} SFR for
NGC~4395 that follows from using the far-infrared luminosity described in \citet{LR00}, who
used the calibration by \citet{buatxu96}. This highlights the difficulty in
determining extragalactic star formation rates from integrated light.

\subsection{Ages and masses of clusters}

After identifying the star cluster candidates (see Sect. 4) in the two fields,
we need to estimate their masses and ages. These parameters were
estimated using the program AnalySED, created by
\citet{anders04}. This program determines the masses and ages
using GALEV SSP evolutionary models \citep{schulz02}. 
The parameters used for AnalySED are a Salpeter IMF \citep{salpeter55} in the mass range
0.10 to 100 M$_\odot$, Padova isochrones \citep{girardi00}, and a LMC metallicity.
AnalySED performs a match, comparing the observed and
theoretical spectral energy distributions (SEDs), based on GALEV SSP models. As
output, AnalySED provides, among other parameters, an estimation of the
mass and the age for each cluster.

In addition to our magnitude limit of $V=23$ ($M_V\sim-5$) for our clusters, we also define a lower mass
limit at $10^{2.5}$ M$_{\sun}$ to increase the likelihood that extended objects are true clusters 
and not just chance projections of a few bright stars. However, over most of the age range ($Log(age)\ge7$), 
these false detections would fall below our magnitude limit.

\begin{figure}[!t]
	\centering
	\includegraphics[width=\columnwidth]{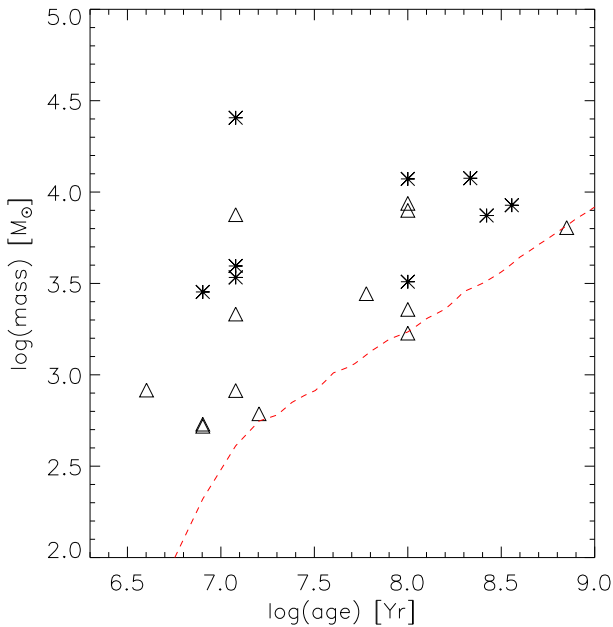}
	\caption{Age-mass distribution of the clusters detected in our two
		      fields observed. The red dashed line represents M$_V=-5$.
		      Symbols are the same as in Fig. \ref{fig:twocolor}.}
	\label{fig:agemas}
\end{figure}

Figure~\ref{fig:agemas} shows the age-mass diagram of the cluster candidates in 
NGC~4395. The number of clusters detected is rather small because the two HST 
pointings do not cover the whole galaxy and our coverage for clusters is reduced further 
by our requirement of the WFPC2 pointings for age-dating.
We see that neither massive ($M\geq10^5$ M$_{\odot}$) nor old ($\tau\ge10^9$ yr) clusters
are detected, in agreement with previous work \citep{LR99,mora09}. We identified
few clusters with $\tau>100$ Myr for LMC metallicity in agreement with \citet{mora09},
although the number of objects was greater in their work. Nevertheless, the number of objects detected at these ages (and greater) varies,
as can be see in Fig. 7 of Mora et al., depending on the metallicity used to estimate the parameters.


\section {Summary, discussion, and conclusions}
We have described our procedures to obtain the age distributions
of field stars and star clusters in HST images of nearby ($D\approx4$ Mpc)
galaxies, in particular, we have described our implementation of the
synthetic CMD method and tests of our program developed for this
purpose. We found that the code recovers the star formation histories of synthetic
populations with good accuracy, while errors in the derived SFHs
of true populations are dominated by systematics errors. 
We have derived the SFH for NGC~4395 for three different metallicities
and find an approximately constant SFR over the past few hundred Myr
assuming LMC-like metallicity. Only a modest amount of internal extinction ($A_B\approx0.08$) 
in NGC~4395 is required in addition to the Galactic foreground extinction 
to match the data. Hence, uncertainties in the total extinction
is not a major source of error compared to, for example, binary stars and the choice of isochrones.
Different assumptions may lead to changes in
our estimated SFRs of up to a factor of 2--4 in specific age bins, although
the global average will be less affected. Poissonian errors do not contribute substantially 
to the uncertainties because of the large number of field stars used in our analysis.

After estimating the SFH of the field stars and both the age and mass distributions
of star clusters, we can now measure which fraction of stars are in clusters.
We compare the fractions for cluster ages between $10^7$ and $10^8$
years, since disruption due to tidal shocks (GMC and spiral arm encounters) and evaporation is expected to become significant at older ages,
while younger objects may still be unbound and prone to ``infant mortality''.

A rough estimate of the cluster formation rate (CFR) can be obtained by dividing the total amount of mass in
clusters observed by the age range. From $10^7\leq\tau\leq10^8$ yr, the total mass
observed in Fig. \ref{fig:agemas} is $M_{obs}\sim8.3\times10^4$ M$_\odot$, which
infers a $\log(CFR)\approx-3.03$ [M$_\odot$yr$^{-1}$]. We correct this for the smaller
area covered by the WFPC2 camera compared to the ACS (a factor of $\sim2.27$) and
thus obtain $\log(CFR)\approx-2.68$ [M$_\odot$Yr$^{-1}$] for the same equivalent area covered
by our field star data.

A magnitude-limited sample does not allow us to observe all the 
clusters in the age range. Using the observed total mass and assuming a cluster
mass function ($\Psi(m)$), we can estimate the total mass in clusters for a certain age
range. \citet{larsen09} showed that a Schechter function (with $M_\star=2\times10^5$M$_\odot$) 
can model the initial cluster mass function in present-day spiral discs. However, the 
canonical cluster mass function, a power law with index $-2$, can be used for the same 
purpose. To estimate the cluster formation rate we used both functions and calculated
the total mass in the age range mentioned above. Since the most massive clusters
observed have M$< 10^5$ M$_\odot$, it makes little difference whether we adopt
the Schechter function or an untruncated power-law.

The total mass in the cluster system was estimated using the equation

\begin{equation}
M_{tot}=M_{obs}\times
\frac
{ (t_2-t_1) \int_{m_{lo}}^{m_{up}}m\Psi(m)\,dm} 
{ \int_{t_1}^{t_2} \int_{m_{lim(t)}}^{m_{up}}m\Psi(m)\,dm\,dt }
\ ,
\label{equ:mtotclu}
\end{equation}
where $t_1,t_2$ represent the time interval ($t_1 = 10^7$ years, $t_2 = 10^8$ years), 
$m_{lo}$ and $m_{up}$ are the lower and upper mass limits over which the mass is 
normalized ($m_{lo}=10$ and $m_{up}=10^6$ M$_\odot$), and $m_{lim(t)}$ is the (age-dependent) detection limit of our cluster 
sample (dashed line in Fig. \ref{fig:agemas}). Since the last parameter, $m_{lim}$, is derived from the SSP models used,
this integral must be evaluated numerically. Equation (\ref{equ:mtotclu}) assumes a uniform
cluster age distribution, which is the most conservative choice we can make 
given the small number of clusters in our sample and the uncertainties in the age 
determinations (Fig.~\ref{fig:twocolor}). We also assume that $\Psi(m)$ is age-independent,
meaning that we ignore disruption. Applying the correction for undetected clusters,
we obtained values for the CFR of $\log(CFR)=-2.37$ and $\log(CFR)=-2.35$ for power law and Schechter functions
respectively, values that are corrected for the area covered by the HST detectors mentioned above.
We have also tried applying Eq.~(\ref{equ:mtotclu}) to 3 sub-bins 
in age, which changed the total overall CFR by a very small amount. We note that, strictly speaking, we are not measuring 
the true cluster \emph{formation} rate, but rather the current \emph{age} distribution of surviving clusters. Dividing these
numbers by the SFR ($SFR=0.17$ M$_{\odot}$Yr$^{-1}$), we obtain a ratio of $\Gamma = CFR/SFR=0.0263$.

\begin{figure}[!t]
	\centering
	\includegraphics[width=\columnwidth]{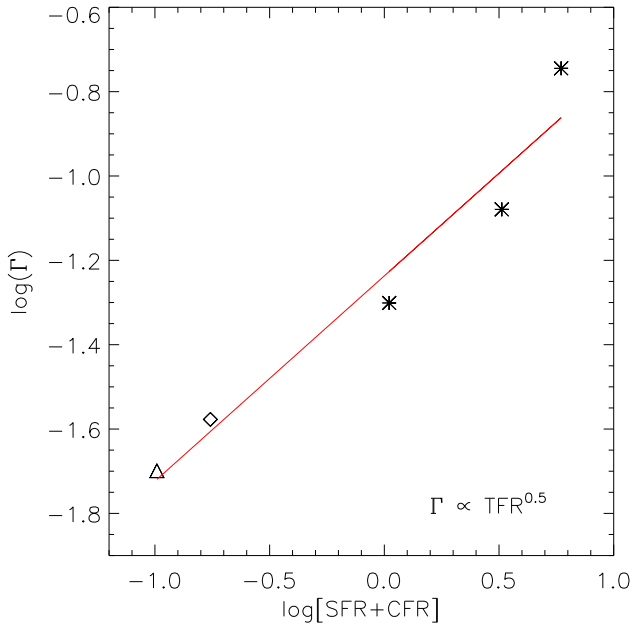}
	\caption{Relation between the cluster formation efficiency ($\Gamma$) and the total formation rate ($TFR=SFR+CFR$).
		     An apparent power law function fits the date following the equation $\Gamma \approx TFR^{0.5}$. Symbols represent:
		     $\star$ are data from \citet{gieles09} for the galaxies M74, M101 and M51; $\triangle$ is for the SMC 
		     \citep{gielesbastian08}; and $\diamond$ our estimation for NGC 4395.}
	\label{fig:gammarel}
\end{figure}

Our estimate of $\Gamma$ is lower than the value $\Gamma\sim0.08$ obtained by
\citet{bastian08}. \citet{gieles09} estimated $\Gamma$ for three galaxies namely M74, M101, and M51. The
cluster formation rate was estimated by comparing theoretical
with empirical luminosity functions. Gieles suggested that $\Gamma$ has the 
tendency to increase with SFR. \citet{gielesbastian08} estimated
$\Gamma\approx0.02$ for the SMC using an accurate determination of the mean 
global star formation rate based on the field stars (SFR$\approx0.1$ M$_\odot$Yr$^{-1}$)
performed by \citet{harriszaritsky04} and the CFR calculated in their work. Other studies have found
the same value for $\Gamma$ in the SMC \citep[see ][]{goddard10}.
We took the data from \citet{gieles09} (see Table 1 in his article), and \citet{gielesbastian08}
and plotted the total formation rate ($TFR=SFR+CFR$) against $\Gamma$
for these four galaxies and included our CFR and SFR for NGC 4395 
to attempt to detect any correlation. Figure \ref{fig:gammarel} shows the plot of $\log(\Gamma)$ versus $\log(TFR)$. There is an apparent
correlation between these two values, which can be approximated by a power law
$\Gamma \propto TFR^{0.5}$. Although only five points are plotted and the TFR in NGC 4395 does not cover the 
whole galaxy, Fig.~\ref{fig:gammarel} does suggest that the relatively low $\Gamma$ value
we derive for NGC 4395 is consistent with a general trend. 
In a future study (Silva-Villa et al.\ 2010, in prep.), we will include
more points in this plot for galaxies spanning a range in TFR 
and check the validity of this result.

\citet{goddard10} found an empirical relation between $\Gamma$ and the SFR density 
($\Sigma_{SFR}$ [M$_{\odot}$yr$^{-1}$Kpc$^{-2}$]). We estimated 
$\Sigma_{SFR}$ to be $\sim4.65\times10^{-3}$ M$_{\odot}$Yr$^{-1}$Kpc$^{-2}$ for NGC~4395, which leads to $\Gamma\approx8\%$
based on the \citet{goddard10} relation. This value is higher than
that estimated in this work $\Gamma\approx3\%$. We point out that the area covered 
by our observations is not the total area of the galaxy and so the SFR
density found in this work might be higher than the global average. 


\begin{acknowledgements}
  Silva-Villa would like to thank Peter Anders for help with 
  the program AnalySED. We thank C. U. Keller and the referee for comments that 
  helped improve this article.
\end{acknowledgements}

\bibliographystyle{aa}
\bibliography{articlengc4395}

\begin{thebibliography}{66}
\expandafter\ifx\csname natexlab\endcsname\relax\def\natexlab#1{#1}\fi

\bibitem[{{Anders} {et~al.}(2004){Anders}, {Bissantz}, {Fritze-v.~Alvensleben},
  \& {de Grijs}}]{anders04}
{Anders}, P., {Bissantz}, N., {Fritze-v.~Alvensleben}, U., \& {de Grijs}, R.
  2004, \mnras, 347, 196

\bibitem[{{Anders} \& {Fritze-v.~Alvensleben}(2003)}]{andersfritze03}
{Anders}, P. \& {Fritze-v.~Alvensleben}, U. 2003, \aap, 401, 1063

\bibitem[{{Annibali} {et~al.}(2009){Annibali}, {Tosi}, {Monelli}, {Sirianni},
  {Montegriffo}, {Aloisi}, \& {Greggio}}]{annibali09}
{Annibali}, F., {Tosi}, M., {Monelli}, M., {et~al.} 2009, \aj, 138, 169

\bibitem[{{Aparicio} \& {Gallart}(2004)}]{apariciogallart04}
{Aparicio}, A. \& {Gallart}, C. 2004, \aj, 128, 1465

\bibitem[{{Aparicio} {et~al.}(1997){Aparicio}, {Gallart}, \&
  {Bertelli}}]{aparicio97}
{Aparicio}, A., {Gallart}, C., \& {Bertelli}, G. 1997, \aj, 114, 669

\bibitem[{{Aparicio} \& {Hidalgo}(2009)}]{apariciohidalgo09}
{Aparicio}, A. \& {Hidalgo}, S.~L. 2009, \aj, 138, 558

\bibitem[{{Barker} {et~al.}(2007){Barker}, {Sarajedini}, {Geisler}, {Harding},
  \& {Schommer}}]{barker07}
{Barker}, M.~K., {Sarajedini}, A., {Geisler}, D., {Harding}, P., \& {Schommer},
  R. 2007, \aj, 133, 1138

\bibitem[{{Bastian}(2008)}]{bastian08}
{Bastian}, N. 2008, \mnras, 390, 759

\bibitem[{{Bastian} \& {Gieles}(2008)}]{bastiangieles08}
{Bastian}, N. \& {Gieles}, M. 2008, in Astronomical Society of the Pacific
  Conference Series, Vol. 388, Mass Loss from Stars and the Evolution of
  Stellar Clusters, ed. {A.~de Koter, L.~J.~Smith, \& L.~B.~F.~M.~Waters}, 353

\bibitem[{{Baumgardt}(2009)}]{baumgardt09}
{Baumgardt}, H. 2009, {Dissolution of Globular Clusters}, ed. {Richtler, T.~\&
  Larsen, S.}, 387

\bibitem[{{Bertelli} {et~al.}(1994){Bertelli}, {Bressan}, {Chiosi}, {Fagotto},
  \& {Nasi}}]{bertelli94}
{Bertelli}, G., {Bressan}, A., {Chiosi}, C., {Fagotto}, F., \& {Nasi}, E. 1994,
  \aaps, 106, 275

\bibitem[{{Bertin} \& {Arnouts}(1996)}]{bertinarnouts96}
{Bertin}, E. \& {Arnouts}, S. 1996, \aaps, 117, 393

\bibitem[{{Boutloukos} \& {Lamers}(2003)}]{BL03}
{Boutloukos}, S.~G. \& {Lamers}, H.~J.~G.~L.~M. 2003, \mnras, 338, 717

\bibitem[{{Brown} {et~al.}(2008){Brown}, {Beaton}, {Chiba}, {Ferguson},
  {Gilbert}, {Guhathakurta}, {Iye}, {Kalirai}, {Koch}, {Komiyama}, {Majewski},
  {Reitzel}, {Renzini}, {Rich}, {Smith}, {Sweigart}, \& {Tanaka}}]{brown08}
{Brown}, T.~M., {Beaton}, R., {Chiba}, M., {et~al.} 2008, \apjl, 685, L121

\bibitem[{{Buat} \& {Xu}(1996)}]{buatxu96}
{Buat}, V. \& {Xu}, C. 1996, \aap, 306, 61

\bibitem[{{Cervi{\~n}o} \& {Luridiana}(2006)}]{cervignoluridiana06}
{Cervi{\~n}o}, M. \& {Luridiana}, V. 2006, \aap, 451, 475

\bibitem[{{Cignoni} \& {Tosi}(2010)}]{cignonitosi09}
{Cignoni}, M. \& {Tosi}, M. 2010, Advances in Astronomy, 2010, 3

\bibitem[{{Cole} {et~al.}(2007){Cole}, {Skillman}, {Tolstoy}, {Gallagher},
  {Aparicio}, {Dolphin}, {Gallart}, {Hidalgo}, {Saha}, {Stetson}, \&
  {Weisz}}]{cole07}
{Cole}, A.~A., {Skillman}, E.~D., {Tolstoy}, E., {et~al.} 2007, \apjl, 659, L17

\bibitem[{{Dolphin}(1997)}]{dolphin97}
{Dolphin}, A. 1997, New Astronomy, 2, 397

\bibitem[{{Dolphin}(2000)}]{dolphin00}
{Dolphin}, A.~E. 2000, \pasp, 112, 1397

\bibitem[{{Edvardsson} {et~al.}(1993){Edvardsson}, {Andersen}, {Gustafsson},
  {Lambert}, {Nissen}, \& {Tomkin}}]{edvardsson93}
{Edvardsson}, B., {Andersen}, J., {Gustafsson}, B., {et~al.} 1993, \aap, 275,
  101

\bibitem[{{Elmegreen}(2008)}]{elmegreen08}
{Elmegreen}, B.~G. 2008, 388, 249

\bibitem[{{Fall}(2006)}]{fall06}
{Fall}, S.~M. 2006, \apj, 652, 1129

\bibitem[{{Filippenko} \& {Sargent}(1989)}]{filippenkosargent89}
{Filippenko}, A.~V. \& {Sargent}, W.~L.~W. 1989, \apjl, 342, L11

\bibitem[{{Forbes} {et~al.}(1997){Forbes}, {Brodie}, \& {Huchra}}]{forbes97}
{Forbes}, D.~A., {Brodie}, J.~P., \& {Huchra}, J. 1997, \aj, 113, 887

\bibitem[{{Gieles}(2009)}]{gieles09}
{Gieles}, M. 2009, ArXiv e-prints

\bibitem[{{Gieles} \& {Bastian}(2008)}]{gielesbastian08}
{Gieles}, M. \& {Bastian}, N. 2008, \aap, 482, 165

\bibitem[{{Girardi} {et~al.}(2000){Girardi}, {Bressan}, {Bertelli}, \&
  {Chiosi}}]{girardi00}
{Girardi}, L., {Bressan}, A., {Bertelli}, G., \& {Chiosi}, C. 2000, \aaps, 141,
  371

\bibitem[{{Girardi} {et~al.}(1995){Girardi}, {Chiosi}, {Bertelli}, \&
  {Bressan}}]{girardi95}
{Girardi}, L., {Chiosi}, C., {Bertelli}, G., \& {Bressan}, A. 1995, \aap, 298,
  87

\bibitem[{{Goddard} {et~al.}(2010){Goddard}, {Bastian}, \&
  {Kennicutt}}]{goddard10}
{Goddard}, Q.~E., {Bastian}, N., \& {Kennicutt}, R.~C. 2010, ArXiv e-prints

\bibitem[{{Grevesse} \& {Sauval}(1998)}]{grevessesauval98}
{Grevesse}, N. \& {Sauval}, A.~J. 1998, Space Science Reviews, 85, 161

\bibitem[{{Harris} \& {Zaritsky}(2001)}]{harriszaritsky01}
{Harris}, J. \& {Zaritsky}, D. 2001, \apjs, 136, 25

\bibitem[{{Harris} \& {Zaritsky}(2002)}]{harriszaritsky02}
{Harris}, J. \& {Zaritsky}, D. 2002, in Astronomical Society of the Pacific
  Conference Series, Vol. 274, Observed HR Diagrams and Stellar Evolution, ed.
  {T.~Lejeune \& J.~Fernandes}, 600

\bibitem[{{Harris} \& {Zaritsky}(2004)}]{harriszaritsky04}
{Harris}, J. \& {Zaritsky}, D. 2004, \aj, 127, 1531

\bibitem[{{Harris} \& {Zaritsky}(2009)}]{harriszaritsky09}
{Harris}, J. \& {Zaritsky}, D. 2009, \aj, 138, 1243

\bibitem[{{King}(1962)}]{king62}
{King}, I. 1962, \aj, 67, 471

\bibitem[{{Koekemoer} {et~al.}(2002){Koekemoer}, {Fruchter}, {Hook}, \&
  {Hack}}]{koekemoer02}
{Koekemoer}, A.~M., {Fruchter}, A.~S., {Hook}, R.~N., \& {Hack}, W. 2002, in
  The 2002 HST Calibration Workshop : Hubble after the Installation of the ACS
  and the NICMOS Cooling System, ed. {S.~Arribas, A.~Koekemoer, \&
  B.~Whitmore}, 337

\bibitem[{{Kruijssen} \& {Lamers}(2008)}]{kruijssenlamers08}
{Kruijssen}, J.~M.~D. \& {Lamers}, H.~J.~G.~L.~M. 2008, in Astronomical Society
  of the Pacific Conference Series, Vol. 396, Astronomical Society of the
  Pacific Conference Series, ed. {J.~G.~Funes \& E.~M.~Corsini}, 149--+

\bibitem[{{Lada} \& {Lada}(2003)}]{LL03}
{Lada}, C.~J. \& {Lada}, E.~A. 2003, \araa, 41, 57

\bibitem[{{Lamers} \& {Gieles}(2008)}]{lamersgieles08}
{Lamers}, H.~J.~G.~L.~M. \& {Gieles}, M. 2008, in Astronomical Society of the
  Pacific Conference Series, Vol. 388, Mass Loss from Stars and the Evolution
  of Stellar Clusters, ed. {A.~de Koter, L.~J.~Smith, \& L.~B.~F.~M.~Waters},
  367

\bibitem[{{Lamers} {et~al.}(2005){Lamers}, {Gieles}, {Bastian}, {Baumgardt},
  {Kharchenko}, \& {Portegies Zwart}}]{lamers05}
{Lamers}, H.~J.~G.~L.~M., {Gieles}, M., {Bastian}, N., {et~al.} 2005, \aap,
  441, 117

\bibitem[{{Larsen}(1999)}]{larsen99}
{Larsen}, S.~S. 1999, \aaps, 139, 393

\bibitem[{{Larsen}(2009)}]{larsen09}
{Larsen}, S.~S. 2009, \aap, 503, 467

\bibitem[{{Larsen} {et~al.}(2002){Larsen}, {Efremov}, {Elmegreen}, {Alfaro},
  {Battinelli}, {Hodge}, \& {Richtler}}]{larsen02}
{Larsen}, S.~S., {Efremov}, Y.~N., {Elmegreen}, B.~G., {et~al.} 2002, \apj,
  567, 896

\bibitem[{{Larsen} {et~al.}(2001){Larsen}, {Forbes}, \& {Brodie}}]{larsen01}
{Larsen}, S.~S., {Forbes}, D.~A., \& {Brodie}, J.~P. 2001, \mnras, 327, 1116

\bibitem[{{Larsen} {et~al.}(2007){Larsen}, {Mora}, {Brodie}, \&
  {Richtler}}]{larsen07}
{Larsen}, S.~S., {Mora}, M.~D., {Brodie}, J.~P., \& {Richtler}, T. 2007, in IAU
  Symposium, Vol. 241, IAU Symposium, ed. {A.~Vazdekis \& R.~F.~Peletier},
  435--439

\bibitem[{{Larsen} \& {Richtler}(1999)}]{LR99}
{Larsen}, S.~S. \& {Richtler}, T. 1999, \aap, 345, 59

\bibitem[{{Larsen} \& {Richtler}(2000)}]{LR00}
{Larsen}, S.~S. \& {Richtler}, T. 2000, \aap, 354, 836

\bibitem[{{Lejeune} \& {Schaerer}(2001)}]{lejueneschaerer01}
{Lejeune}, T. \& {Schaerer}, D. 2001, \aap, 366, 538

\bibitem[{{Ma{\'{\i}}z Apell{\'a}niz}(2009)}]{maizapellaniz09}
{Ma{\'{\i}}z Apell{\'a}niz}, J. 2009, \apj, 699, 1938

\bibitem[{{Marigo} {et~al.}(2008){Marigo}, {Girardi}, {Bressan}, {Groenewegen},
  {Silva}, \& {Granato}}]{marigo08}
{Marigo}, P., {Girardi}, L., {Bressan}, A., {et~al.} 2008, \aap, 482, 883

\bibitem[{{Mateo}(1998)}]{mateo98}
{Mateo}, M.~L. 1998, \araa, 36, 435

\bibitem[{{Mora} {et~al.}(2007){Mora}, {Larsen}, \& {Kissler-Patig}}]{mora07}
{Mora}, M.~D., {Larsen}, S.~S., \& {Kissler-Patig}, M. 2007, \aap, 464, 495

\bibitem[{{Mora} {et~al.}(2009){Mora}, {Larsen}, {Kissler-Patig}, {Brodie}, \&
  {Richtler}}]{mora09}
{Mora}, M.~D., {Larsen}, S.~S., {Kissler-Patig}, M., {Brodie}, J.~P., \&
  {Richtler}, T. 2009, \aap, 501, 949

\bibitem[{{Rejkuba} {et~al.}(2004){Rejkuba}, {Greggio}, \&
  {Zoccali}}]{rejkuba04}
{Rejkuba}, M., {Greggio}, L., \& {Zoccali}, M. 2004, \aap, 415, 915

\bibitem[{{Roy} {et~al.}(1996){Roy}, {Belley}, {Dutil}, \& {Martin}}]{roy96}
{Roy}, J., {Belley}, J., {Dutil}, Y., \& {Martin}, P. 1996, \apj, 460, 284

\bibitem[{{Salpeter}(1955)}]{salpeter55}
{Salpeter}, E.~E. 1955, \apj, 121, 161

\bibitem[{{Schlegel} {et~al.}(1998){Schlegel}, {Finkbeiner}, \&
  {Davis}}]{schlegel98}
{Schlegel}, D.~J., {Finkbeiner}, D.~P., \& {Davis}, M. 1998, \apj, 500, 525

\bibitem[{{Schulz} {et~al.}(2002){Schulz}, {Fritze-v.~Alvensleben},
  {M{\"o}ller}, \& {Fricke}}]{schulz02}
{Schulz}, J., {Fritze-v.~Alvensleben}, U., {M{\"o}ller}, C.~S., \& {Fricke},
  K.~J. 2002, \aap, 392, 1

\bibitem[{{Sirianni} {et~al.}(2005){Sirianni}, {Jee}, {Ben{\'{\i}}tez},
  {Blakeslee}, {Martel}, {Meurer}, {Clampin}, {De Marchi}, {Ford}, {Gilliland},
  {Hartig}, {Illingworth}, {Mack}, \& {McCann}}]{sirianni05}
{Sirianni}, M., {Jee}, M.~J., {Ben{\'{\i}}tez}, N., {et~al.} 2005, \pasp, 117,
  1049

\bibitem[{{Skillman} \& {Gallart}(2002)}]{coimbra02}
{Skillman}, E.~D. \& {Gallart}, C. 2002, in Astronomical Society of the Pacific
  Conference Series, Vol. 274, Observed HR Diagrams and Stellar Evolution, ed.
  {T.~Lejeune \& J.~Fernandes}, 535

\bibitem[{{Spitzer}(1987)}]{spitzer87}
{Spitzer}, L. 1987, {Dynamical evolution of globular clusters}, ed. L.~Spitzer

\bibitem[{{Tosi} {et~al.}(1991){Tosi}, {Greggio}, {Marconi}, \&
  {Focardi}}]{tosi91}
{Tosi}, M., {Greggio}, L., {Marconi}, G., \& {Focardi}, P. 1991, \aj, 102, 951

\bibitem[{{Whitmore}(2007)}]{whitmore07}
{Whitmore}, B.~C. 2007, in IAU Symposium, Vol. 237, IAU Symposium, ed.
  {B.~G.~Elmegreen \& J.~Palous}, 222--229

\bibitem[{{Williams} {et~al.}(2009){Williams}, {Dalcanton}, {Seth}, {Weisz},
  {Dolphin}, {Skillman}, {Harris}, {Holtzman}, {Girardi}, {de Jong}, {Olsen},
  {Cole}, {Gallart}, {Gogarten}, {Hidalgo}, {Mateo}, {Rosema}, {Stetson}, \&
  {Quinn}}]{williams09}
{Williams}, B.~F., {Dalcanton}, J.~J., {Seth}, A.~C., {et~al.} 2009, \aj, 137,
  419

\bibitem[{{Young} {et~al.}(2007){Young}, {Skillman}, {Weisz}, \&
  {Dolphin}}]{young07}
{Young}, L.~M., {Skillman}, E.~D., {Weisz}, D.~R., \& {Dolphin}, A.~E. 2007,
  \apj, 659, 331

\end{thebibliography}
\end{document}